\documentclass[12pt]{article}

\usepackage{epstopdf,amsfonts,amsmath,amssymb}

\makeatletter
\@addtoreset{equation}{section}
\makeatother
\newtheorem{theorem}{Theorem}[section]

\newtheorem{remark}{Remark}[section]
\newtheorem{proposition}{Proposition}[section]
\newtheorem{lemma}{Lemma}[section]
\newtheorem{corollary}{Corollary}[section]
\newtheorem{definition}{Definition}[section]

\newtheorem{example}[theorem]{Example}

\def\br{\begin{remark}\rm\small}
\def\er{\end{remark}}
\def\bt{\begin{theorem}}
\def\et{\end{theorem}}
\def\bd{\begin{definition}}
\def\ed{\end{definition}}
\def\bp{\begin{proposition}}
\def\ep{\end{proposition}}
\def\bl{\begin{lemma}}
\def\el{\end{lemma}}
\def\bc{\begin{corollary}}
\def\ec{\end{corollary}}
\def\bex{\begin{example}}
\def\eex{\end{example}}

\def\beaq{\begin{eqnarray}}
\def\eeaq{\end{eqnarray}}

\newcommand{\be}{\begin{equation}}
\newcommand{\beq}{\begin{equation}}
\newcommand{\eeq}{\end{equation}}
\newcommand{\bea}{\begin{eqnarray}}
\newcommand{\eea}{\end{eqnarray}}

\newcommand{\Tr}{{\operatorname{Tr}}}

\newcommand{\CC}{{\mathbb C}}
\newcommand{\RR}{{\mathbb R}}
\newcommand{\ZZ}{{\mathbb Z}}

\newcommand{\curve}{{\Sigma}}

\newcommand{\curverond}{\overset{\circ}{\curve}}

\newcommand{\Newt}{{\mathcal{N}}}

\newcommand{\acycle}{{\cal A}}

\newcommand{\modsp}{{\mathcal M}}

\newcommand{\binomial}[2]{\begin{pmatrix}#1 \cr #2 \end{pmatrix}}

\newcommand{\Field}{{\mathbb F}}

\newcommand{\sn}{\operatorname{sn}}
\newcommand{\cn}{\operatorname{cn}}
\newcommand{\dn}{\operatorname{dn}}

\newcommand{\y}{{\rm y}}

\newcommand{\td}{\tilde}
\usepackage[style=alphabetic,maxalphanames=4,giveninits,sorting=nyt,%
maxbibnames=99,backend=bibtex]{biblatex}
\renewbibmacro{in:}{}
\usepackage{csquotes}

\newcommand{\Res}{\mathop{\,\rm Res\,}}
\usepackage[pdftex]{hyperref}
\hypersetup{colorlinks, urlcolor=magenta, citecolor=red, linkcolor=blue, filecolor=black}

\begin{document}

\sloppy

\pagestyle{empty}
\addtolength{\baselineskip}{0.20\baselineskip}
\begin{center}
\vspace{26pt}
{\large \bf {Topological Recursion, explicit ABCD tensors and implementation}}
\vspace{26pt}

{\sl B.\ Eynard}\footnote{Université Paris-Saclay, CNRS, CEA, Institut de physique théorique, 91191, Gif-sur-Yvette, France.}
\footnote{CRM, Centre de recherches math\'ematiques  de Montr\'eal, Montr\'eal, QC, Canada.}

\end{center}
\vspace{20pt}
\begin{center}
{\bf Abstract}

We provide explicit expressions of ABCD tensors for the most classical classes of spectral curves.
And we discuss  algorithmic implementation of Topological Recursion.

\end{center}

%


\vspace{26pt}
\pagestyle{plain}
\setcounter{page}{1}


\tableofcontents


\section{Introduction}

{\it Topological Recursion} was invented in order to compute recursively the large size asymptotic expansion of random matrices \cite{Ey04}, and then it was realized that Topological Recursion is ubiquitous in mathematical physics, beyond random matrices, it has applications to enumerative geometry, string theory, integrable systems, conformal field theory, combinatorics of maps, statistical physics...etc.

In \cite{EO07} Topological Recursion was formulated as a process that to a spectral curve (a Riemann surface with extra structure) associates a sequence of differential forms, called the {\it invariants} of the spectral curve.
These invariants often coincide with classical geometric invariants. 
For example:

\medskip

\begin{tabular}{|l|c|l|}
\hline
    spectral curve & & $\to$ T.R. invariants \\
\hline
     Airy & $y=\sqrt{x}$  & Witten-Kontsevich intersection numbers  \\
     Sine & $y= \sin{(2\pi z)}$ & Weil-Petersson volumes \\
     Mirror of toric CY3-fold & $P(e^x,e^y)=0$ & Gromov-Witten invariants \\
     semi-circle (GUE) & $y=\sqrt{4-x^2}$  & higher Catalan numbers \\
     A-polynomial of knot & $P(e^x,e^y)=0$ & Jones polynomial \\
     classical stress--energy  & $y^2=T(x)$ & CFT conformal blocs \\
     \hspace{15mm} tensor &  &  \\
     etc... & & \\
\hline
\end{tabular}

\medskip

There exists many lectures, introductory lecture notes and review articles on topological recursion and its applications.

\medskip

In \cite{KS2017}, Topological Recursion was reformulated as "\textbf{Quantum Airy Structure}"\index{Quantum Airy Structure}, in terms of 4 tensors called $A,B,C,D$, which was further developed in \cite{ABCD_2024}.

The ABCD tensor formulation amounts to decomposing the invariants on a basis of forms, and consider the recursion for the coefficients.
This is how Topological recursion is actually implemented algorithmically.

\medskip

In this article, we give explicit expressions of the ABCD tensors for many classical spectral curves, in a way used for actual implementation in a python package.
We also discuss the algorithmic issues and implementation.

\section{Topological Recursion and ABCD tensors }

Consider a Riemann surface $\curverond$ called the base curve.
Most often $\curverond=\CC P^1 = \CC\cup\{\infty\}$ the Riemann sphere, and this is what we shall consider in all examples below.

\subsection{Spectral curve}
A \textbf{spectral curve}\index{spectral curve} is the following data
\beq
\left\{
\begin{array}{l}
\curve = \text{Riemann surface} \cr
x: \curve \to \curverond \quad = \text{ramified covering map} \cr
\y = ydx = \text{ meromorphic 1-form on } \curve   \cr
B(z_1,z_2) = \text{fundamental 2nd kind differential }, B\in H^0(\curve\times\curve, K_\curve\overset{\text{Sym}}{\boxtimes} K_\curve)_{\Res=1} \cr
R = \{\text{ramification points}\} \cr
a\in R \ : \ \ 
\sigma_a, \ \text{local anayltic involution such that } x(\sigma_a(z))=x(z)  \text{ and } \sigma_a(a)=a.
\end{array}
\right.
\eeq
For regular spectral curves, ramification points are the zeros of $dx$, and we assume that they are simple zeros.
Therefore near $z=a$, $x$ is $2:1$, and there exists a unique local involution  $\sigma_a\neq \text{Id}$, such that $\sigma_a(a)=a$ and $x(\sigma_a(z))=x(z)$ in a neighborhood of $a$.

We shall consider non-regular spectral curves in section \ref{sec:higherbp} below.

\smallskip 

Then we define the \textbf{recursion kernel}\index{recursion kernel}:
\beq
K_a(z_1,z) = \frac{dS(z_1,z)}{(y(z)-y(\sigma_a(z)))dx(z)}
\quad \text{where} \quad
 dS(z_1,z) = \frac12 \int_{\sigma_a(z)}^z B(z_1,.) 
\eeq
which is meromorphic for $z$ in a neighborhood of the ramification point $a$, with a simple pole at $z=a$.

\subsection{TR invariants}

\textbf{Topological Recursion}\index{Topological Recursion} is a recursive procedure that to a spectral curve, associates its \textbf{invariants}\index{invariants} $\omega_{g,n}$ defined as follows:
\bea
\omega_{0,1} &=& ydx \cr
\omega_{0,2} &=& B \cr
\omega_{0,3}(z_1,z_2,z_3) &=& \sum_{a\in R} \Res_{z\to a} K_a(z_1,z) \left[B(z,z_2)B(\sigma_a(z),z_3) + B(z,z_3)B(\sigma_a(z),z_2)\right] \cr
 &=& -2 \sum_{a\in R} \Res_{z\to a} K_a(z_1,z) B(z,z_2) B(z,z_3) \cr
 &=&  -\sum_{a\in R} \Res_{z\to a} \frac{B(z,z_1) B(z,z_2) B(z,z_3)}{dx(z)dy(z)}  \cr
\omega_{1,1}(z_1) &=& \sum_{a\in R} \Res_{z\to a} K_a(z_1,z) B(z,\sigma_a(z)) 
\eea
and for $n\geq 1$ and $2g-2+n>1$
\bea\label{eq:TopRec}
\omega_{g,n}(z_1,\dots,z_n) 
&=& 2 \sum_{j=2}^{n} \sum_{a\in R} \Res_{z\to a} K_a(z_1,z)  \cr
&& \qquad B(z,z_j) \omega_{g,n}(\sigma_a(z),z_2,\dots, z_{j-1},z_{j+1} , \dots,z_n) \cr
&& + \sum_{a\in R} \Res_{z\to a} K_a(z_1,z) \Big[
\omega_{g-1,n+1}(z,\sigma_a(z),z_2,\dots,z_n) \cr
&& + \sum_{g_1+g_2=g} \sum^{\text{stable}}_{I_1\sqcup I_2 = \{z_2,\dots,z_n\}}
\omega_{g_1,1+|I_1|}(z,I_1) \omega_{g_2,1+|I_2|}(\sigma_a(z),I_2)
\Big]
\eea

For $n=0$, the TR invariants are defined for $g\geq 2$ as
\bea
\mathcal F_g = 
\omega_{g,0}
&=& \frac{1}{2-2g} \sum_{a\in R} \Res_{z\to a} \omega_{g,1}(z) F_{0,1}(z) 
\eea
where $dF_{0,1}=\omega_{0,1}$ (any antiderivative of $\omega_{-0,1}=\y=ydx$ gives the same result).

Here we won't consider $\mathcal F_0$ and $\mathcal F_1$.

\textbf{Homogeneity property}:

If we rescale $ydx \to \lambda ydx$ (i.e. we rescale either $x$ or $y$ or both), then 
\beq\label{eq:homogeneity}
\omega_{g,n}\mapsto \lambda^{2-2g-n} \omega_{g,n}
\eeq
Moreover $\omega_{g,n}$ is invariant under translations $x \mapsto x+c$.

\subsection{Basis of 1-forms}

Consider a vector space $V$, subspace of the space of meromorphic 1-forms on $\curve$, such that for all $2g-2+n>0$ we have 
$$ \omega_{g,n} \in V \otimes \dots \otimes V .$$
$V$ can be chosen as a subspace of  meromorphic 1-forms that have poles at the ramification points,  generated by the coefficients of the Taylor expansion of $B(z,z')$ at $z'\in R$, i.e. by derivatives of $B(z',z)$ with respect to $z'$, at $z'\in R$: 
\beq
V \subset \mathop{\operatorname{Span}}_{a\in R,d\in \ZZ_+}\Big\langle (d/dx(z'))^d (B(z',z)/d\zeta_a(z')) _{z'=a} \Big\rangle
\eeq
(using the local coordinate $\zeta_a(z') = \sqrt{x(z')-x(a)}$ near $a\in R$).
$V$ can be chosen smaller for example when the spectral curve has symmetries.

Consider a basis $\{ d\xi_\alpha \}_{\alpha \in J} $ of $V$.
$\omega_{g,n}$ can be decomposed on the basis:
\bea
\omega_{g,n}(z_1,\dots,z_n) &=& \sum_{\alpha_1;\dots;\alpha_n} F_{g,n}[\alpha_1;\dots;\alpha_n] \prod_{i=1}^n d\xi_{\alpha_i}(z_i) 
\eea
where we assume that the basis is chosen so that the number of terms is finite.
Despite our notation $d\xi_\alpha$ needs not be an exact form.

\textbf{The goal is to be able to compute the coefficients $F_{g,n}[\alpha_1;\dots;\alpha_n]$.}

\smallskip

We also consider the antiderivative (only defined locally):
\bea
\mathcal F_{g,n}(z_1,\dots,z_n) &=& \sum_{\alpha_1;\dots;\alpha_n} F_{g,n}[\alpha_1;\dots;\alpha_n] \prod_{i=1}^n \xi_{\alpha_i}(z_i) .
\eea

Examples of basis:

\begin{itemize}
    \item any basis of 1-forms $d\xi_{a,d}(z)$, such that
$d\xi_{a,d}(z)$ has a pole of degree $2d+2$ at $z=a$, and 
$d\xi_{a,d}(z)+d\xi_{a,d}(\sigma_a(z))$ has no pole at $z=a$.

    \item $d\xi_{a,d}(z) = \Res_{z'\to a} (x(z')-x(a))^{-(d+1/2)} B(z,z')$

    \item $d\xi_{a,0}(z) = \Res_{z'\to a} (x(z')-x(a))^{-1/2} B(z,z')$ and $d\xi_{a,d+1}(z) = d(d\xi_{a,d}(z)/dx(z))$.
    
    \item many other choices are possible...
\end{itemize}

\subsection{TR and ABCD Tensors}

Topological Recursion \eqref{eq:TopRec}, written for the coefficients  amounts to:
\bea
F_{0,3}[\alpha_1;\alpha_2;\alpha_3] &=& A[\alpha_1;\alpha_2;\alpha_3] \cr
F_{1,1}[\alpha_1] &=& D[\alpha_1] 
\eea
and for $2g-2+n>1$:
\bea
 F_{g,n}[\alpha_1;\dots;\alpha_n] 
&=& 2 \sum_{i=2}^n \sum_{\alpha\in J} B[\alpha_1;\alpha_i|\alpha] \ F_{g,n-1}[\alpha;\alpha_2;\dots; \widehat{\alpha_i} ;\dots;\alpha_n] \cr
&&  \sum_{\alpha,\alpha'\in J} C[\alpha_1|\alpha;\alpha'] \ \Big[ F_{g-1,n+1}[\alpha;\alpha';\alpha_2;\dots; \alpha_n] \cr
&& + \sum_{g_1+g_2=g}\sum^{\text{stable}}_{I_1\cup I_2 = \{\alpha_2;\dots;\alpha_n\}}
F_{g_1,1+|I_1|}(\alpha; I_1) F_{g_2,1+|I_2|}(\alpha'; I_2) \Big] \cr
\eea
where we have introduced the following tensors: 

$\bullet$ $A \in V^*\otimes V^* \otimes V^*$:
\bea
&& \sum_{\alpha_1,\alpha_2,\alpha_3}  A[\alpha_1,\alpha_2,\alpha_3] \ d\xi_{\alpha_1}(z_1) d\xi_{\alpha_2}(z_2) d\xi_{\alpha_3}(z_3) \cr
&=& \sum_{a\in R} \Res_{z\to a} \frac{B(z,z_1)B(z,z_2)B(z,z_3)}{dx(z)dy(z)} \cr
&=& -2 \sum_{a\in R} \Res_{z\to a} K_a(z_1,z) B(z,z_1)B(z,z_2) \cr
\eea

$\bullet$ $B \in V^*\otimes V^* \otimes V$:
\bea
\sum_{\alpha_1}\sum_{\alpha_2} B[\alpha_1,\alpha_2|\alpha_3] d\xi_{\alpha_1}(z_1) d\xi_{\alpha_2}(z_2)
&=& \sum_{a\in R} \Res_{z\to a} K_a(z_1,z) B(z,z_2) d\xi_{\alpha_3}(\sigma_a(z)) \cr
&=& - \sum_{a\in R} \Res_{z\to a} K_a(z_1,z) B(z,z_2) d\xi_{\alpha_3}(z) \cr
\eea

$\bullet$ $C \in V^*\otimes V \otimes V$:
\bea
\sum_{\alpha_1} C[\alpha_1|\alpha_2,\alpha_3] d\xi_{\alpha_1}(z_1)
&=& \sum_{a\in R}\Res_{z\to a} K_{a_1}(z_1,z)  d\xi_{\alpha_2}(z) d\xi_{\alpha_3}(\sigma_a(z)) \cr
\eea

$\bullet$ $D \in V^*$:
\bea
\sum_{\alpha_1} D[\alpha_1] d\xi_{\alpha_1}(z_1) &=& \sum_{a\in R}\Res_{z\to a} K_a(z_1,z) B(z,\sigma_a(z)) .
\eea
These tensors are the same as those of Quantum Airy Structures of Kontsevich Soibelman \cite{KS2017} and \cite{ABCD_2024}.
Now we shall compute explicitely $A,B,C,D$ for several examples of spectral curves.


\section{Airy and KdV}

\subsection{Spectral curve}

The \textbf{KdV}\index{KdV} spectral curve with times $\{\td t_k\}_{k\in \ZZ_+}$ is defined as
\beq
\left\{
\begin{array}{l}
\curve=\CC \cr
x=z^2 +\td t_1 \cr
y = -\frac12 \sum_{k\geq 3} \td t_k z^{k-2} \cr
B = \frac{dz_1 dz_2}{(z_1-z_2)^2} \cr
R = \{0\}, \ \ \sigma_0(z)=-z
\end{array}
\right.
\eeq

\textbf{Airy} is the case 
\beq
\td t_k = -2\delta_{k,3}.
\eeq

The \textbf{KP-times} are related to \textbf{KdV-times} as follows:
\bea
t_k = -\delta_{k,1}\td t_1 + \sum_{j\geq 0} \binomial{-k/2}{j} \td t_{k+2j} \td t_1^j.
\eea
In particular if $\td t_1=0$, KdV and KP times are the same $t_k=\td t_k$.

The \textbf{Minimal model} of type $(p,2)$ is such that $\td t_k=0$ for $k>p+2$.

\textbf{Airy} is the minimal model $(1,2)$.

The minimal model $(3,2)$ is often called \textbf{pure-gravity}, or \textbf{Painlevé 1}.

The minimal model $(5,2)$ is often called \textbf{Lee-Yang}.

The case $\td t_{2k+3} = t_{2k+3} =   \frac{(-1)^k (2\pi)^{2k}}{(2k+1)!}$ is called \textbf{Weil-Petersson}.


\subsection{Basis}

We define
\bea
\xi_d(z) = - \frac{(2d-1)!!}{z^{2d+1}}
\quad , \quad
d\xi_d(z) =  \frac{(2d+1)!! dz}{z^{2d+2}} .
\eea

We have
\bea
B(z,z_1) &=& \sum_{k=1}^\infty k z^{k-1}dz \ \frac{dz_1}{z_1^{k+1}} 
\eea
keeping only the odd part under the involution $\sigma_0(z)=-z$, we have
\bea
B_{\text{odd}}(z,z_1) 
&=& \sum_{k=0}^\infty (2k+1) z^{2k}dz \ \frac{dz_1}{z_1^{2k+2}} \cr
&=& \sum_{k=0}^\infty \frac{1}{(2k-1)!!} z^{2k}dz \ d\xi_d(z_1) \cr
\eea
whose antiderivative wrt $z$ is then
\bea
dS_{\text{odd}}(z_1;z) 
&=& \sum_{k=0}^\infty \frac{1}{(2k+1)!!} z^{2k+1} \ d\xi_d(z_1) .
\eea

\subsection{Kernel}

The kernel is
\bea
K_0(z_1,z) 
&=& \frac{dS_{\text{odd}}(z_1;z) }{(y(z)-y(-z))dx(z)} \cr
&=& \frac{-1}{2zdz} \ \left( \sum_{k=0}^\infty \frac{1}{(2k+1)!!} z^{2k} \ d\xi_d(z_1) \right) \left( \sum_{k\geq 0 } \td t_{2k+3} z^{2k} \right)^{-1} \cr
&=& \frac{-1}{2\td t_3}  \ \left( \sum_{k=0}^\infty \frac{1}{(2k+1)!!} z^{2k} \ d\xi_d(z_1) \right) \left( 1+ \sum_{k\geq 1} \frac{\td t_{2k+3}}{\td t_3} z^{2k} \right)^{-1} \ \frac{1}{zdz} \cr
\eea

Let us define:
\bea
T_0 &=& 1 \cr
T_k &=& \sum_{n=1}^k \frac{(-1)^n}{\td t_3^n} \sum_{j_1+\dots+j_n=k, \ j_i>0} \td t_{3+2j_1}\dots  \td t_{3+2j_n}.
\eea
We have
\bea
K_0(z_1,z) 
&=& \frac{-1}{2\td t_3}   \left( \sum_{d=0}^\infty  \frac{d\xi_d(z_1)}{(2d+1)!!} z^{2d}   \right) \left(\sum_{k\geq 0} T_k z^{2k} \right)\ \frac{1}{zdz}  .
\eea

\subsection{First invariants}

We have
\bea
\omega_{0,3}(z_1,z_2,z_3) 
&=& \Res_{z \to 0} K_0(z_1,z) (B(z,z_2)B(-z,z_3)+ B(z,z_3)B(-z,z_2)) \cr
&=& -2 \Res_{z \to 0} K_0(z_1,z) B(z,z_2)B(z,z_3) \cr
&=& \frac{1}{\td t_3} \frac{dz_1}{z_1^2} \Res_{z \to 0} \left(1-z^2/z_1^2 \right)^{-1} \left(1+\sum_{k\geq 1}  T_k z^{2k} \right)\ \frac{1}{zdz} \frac{dz^2 dz_2 dz_3}{(z-z_2)^2(z-z_3)^2} \cr
&=& \frac{1}{\td t_3} \prod_{i=1}^3 \frac{dz_i}{z_i^2} \cr
&=& \frac{1}{\td t_3} \prod_{i=1}^3 d\xi_0(z_i).
\eea

We have
\bea
\omega_{1,1}(z_1) 
&=& \Res_{z \to 0} K_0(z_1,z) B(z,-z) \cr
&=& - \Res_{z \to 0} K_0(z_1,z) \frac{dz^2}{4z^2} \cr
&=& \frac{1}{2\td t_3} \frac{dz_1}{z_1^2}  \Res_{z \to 0} \left(1-z^2/z_1^2 \right)^{-1} \left(1+\sum_{k\geq 1}  T_k z^{2k} \right)\ \frac{1}{zdz}  \frac{dz^2}{4z^2} \cr
&=& \frac{1}{8\td t_3} \frac{dz_1}{z_1^2}  \Res_{z \to 0} \left(1-z^2/z_1^2 \right)^{-1} \left(1+\sum_{k\geq 1}  T_k z^{2k} \right)\ \frac{dz}{z^3}  \cr
&=& \frac{1}{8\td t_3} \frac{dz_1}{z_1^2} \left( \frac{1}{z_1^2} +T_1 \right)  \cr
&=& \frac{1}{8\td t_3}  \left( \frac{1}{3}d\xi_1(z_1) + T_1 d\xi_0(z_1) \right)  \cr
\eea

\subsection{ABCD Tensors}

$\bullet$ $A$
\bea
A[0,0,0] = \frac{1}{\td t_3}
\eea

$\bullet$ $D$
\bea
D[1] = \frac{1}{24\td t_3}
\quad , \quad
D[0] = \frac{-\td t_5}{8\td t_3^2} = \frac{T_1}{8\td t_3} .
\eea

$\bullet$ $B$

\bea
\sum_{d_1}\sum_{d_2} B[d_1,d_2|d_3] d\xi_{d_1}(z_1)d\xi_{d_2}(z_2)
&=&  \Res_{z\to 0} K_0(z_1,z) B(z,z_2) d\xi_{d_3}(-z) \cr
&=& - \Res_{z\to 0} K_0(z_1,z) B(z,z_2) d\xi_{d_3}(z) \cr
\eea

This implies
\bea
 B[d_1,d_2|d_3] 
&=& \frac{1}{2\td t_3 (2d_1+1)!!} \Res_{z\to 0} \left(\sum_{k\geq 0} T_k z^{2k} \right) \frac{z^{2d_1-1}}{dz} \frac{z^{2d_2} dz}{(2d_2-1)!!}  d\xi_{d_3}(z) \cr
&=& \frac{1}{2\td t_3 (2d_1+1)!!} \Res_{z\to 0} \left(\sum_{k\geq 0} T_k z^{2k} \right) \frac{z^{2d_1-1}}{dz} \frac{z^{2d_2} dz}{(2d_2-1)!!}  \frac{(2d_3+1)!! dz}{z^{2d_3+2}} \cr
&=& \frac{(2d_3+1)!!}{2\td t_3 (2d_1+1)!!(2d_2-1)!!} \Res_{z\to 0} \left(\sum_{k\geq 0} T_k z^{2k} \right) z^{2d_1} z^{2d_2} z^{-(2d_3+2)} \frac{dz}{z}  \cr
&=& \frac{(2d_3+1)!!}{2\td t_3 (2d_1+1)!!(2d_2-1)!!} T_{d_3+1-d_1-d_2} \ \delta_{d_1+d_2\leq d_3+1} .
\eea

$\bullet$ $C$

\bea
\sum_{d_1} C[d_1| d_2,d_3] d\xi_{d_1}(z_1)
&=&  \Res_{z\to 0} K_0(z_1,z) d\xi_{d_2}(z) d\xi_{d_3}(-z) \cr
&=& - \Res_{z\to 0} K_0(z_1,z) d\xi_{d_2}(z) d\xi_{d_3}(z) \cr
\eea

This implies
\bea
C[d_1| d_2,d_3] 
&=& \frac{1}{2\td t_3 (2d_1+1)!!} \Res_{z\to 0} \left(\sum_{k\geq 0} T_k z^{2k} \right) \frac{z^{2d_1}}{z dz} \frac{(2d_2+1)!! dz}{z^{2d_2+2}} \frac{(2d_3+1)!! dz}{z^{2d_3+2}} \cr
&=& \frac{(2d_2+1)!!(2d_3+1)!!}{2\td t_3 (2d_1+1)!!} \Res_{z\to 0} \left(\sum_{k\geq 0} T_k z^{2k} \right)z^{2(d_1-d_2-d_3-2)} \frac{dz}{z} \cr
&=& \frac{(2d_2+1)!!(2d_3+1)!!}{2\td t_3 (2d_1+1)!!} T_{d_2+d_3+2-d_1} \ \delta_{d_1\leq d_2+d_3+2} .
\eea

\subsection{Summary KdV}

\subsubsection{General KdV case}

\bea
\omega_{g,n}(z_1,\dots,z_n) &=& \sum_{d_1,\dots,d_n} F_{g,n}[d_1,\dots,d_n] \prod_{i=1}^n d\xi_{d_i}(z_i) \cr
\mathcal F_{g,n}(z_1,\dots,z_n) &=& \sum_{d_1,\dots,d_n} F_{g,n}[d_1,\dots,d_n] \prod_{i=1}^n \xi_{d_i}(z_i)
\eea

\bea
\xi_d(z) = - \frac{(2d-1)!!}{z^{2d+1}}
\quad , \quad
d\xi_d(z) =  \frac{(2d+1)!! dz}{z^{2d+2}} .
\eea

\bea
A[0,0,0] &=& \frac{1}{\td t_3} \cr
D[1] &=& \frac{1}{24\td t_3} 
\quad , \quad
D[0] = \frac{T_1}{8\td t_3} \cr
B[d_1,d_2|d_3] 
&=& \frac{(2d_3+1)!!}{2\td t_3 (2d_1+1)!!(2d_2-1)!!} T_{d_3+1-d_1-d_2} \ \delta_{d_1+d_2\leq d_3+1} \cr
C[d_1| d_2,d_3] 
&=& \frac{(2d_2+1)!!(2d_3+1)!!}{2\td t_3 (2d_1+1)!!} T_{d_2+d_3+2-d_1} \ \delta_{d_1\leq d_2+d_3+2} .
\eea


We recall that for the KdV spectral curve, the invariants are the generating functions of Witten-Kontsevich intersection numbers.

Case $\td t_1=0, \td t_3=-2$:
\bea
F_{g,n}[d_1,\dots,d_n]  
&=& (-2)^{2-2g-n} <\tau_{d_1}\dots \tau_{d_n} e^{\frac12\sum_k (2k-1)!! t_{2k+1} \tau_k } >_g.
\eea
The case $\td t_1=0$ and $\td t_3\neq -2$ can be obtained by the homogeneity \eqref{eq:homogeneity}:
\bea
F_{g,n}[d_1,\dots,d_n]  
&=& \td t_3^{2-2g-n}  <\tau_{d_1}\dots \tau_{d_n} e^{-\frac1{\td t_3}\sum_{k\geq 2} (2k-1)!! \td t_{2k+1} \tau_k } >_g \cr
&=& (-2)^{2-2g-n} <\tau_{d_1}\dots \tau_{d_n} e^{\frac12\sum_{k\geq 1} (2k-1)!! (\td t_{2k+1}+2\delta_{k,1}) \tau_k } >_g .
\eea
The case $\td t_1\neq 0$ is more subttle, see the litterature, for example \cite{EynBook}.

\subsubsection{Summary Airy}
This is the case $T_k=\delta_{k,0}$ and $t_3=-2$:
\beq
\left\{
\begin{array}{l}
x=z^2  \cr
y = z \cr
y^2-x=0
\end{array}
\right.
\eeq
This is also the minimal model $(1,2)$.

\bea
A[0,0,0] &=& \frac{-1}{2} \cr
D[1] = \frac{-1}{48} \cr
B[d_1,d_2|d_3] 
&=& \frac{-1}{4} \frac{(2d_3+1)!!}{(2d_1+1)!!(2d_2-1)!!} \delta_{d_3+1,d_1+d_2} \cr
C[d_1| d_2,d_3] 
&=&  \frac{-1}{4} \frac{(2d_2+1)!!(2d_3+1)!!}{ (2d_1+1)!!} \delta_{d_2+d_3+2,d_1} 
\eea
For the Airy spectral curve, the coefficients $F_{g,n}$ in this basis, are the Witten-Kontsevich intersection numbers
\bea
F_{g,n}[d_1,\dots,d_n]  = (-1)^n 2^{2-2g-n} <\tau_{d_1}\dots \tau_{d_n}>_g.
\eea

This spectral curve is related to the Airy function as follows:
\bea
\psi(z^2) 
&=& z^{-\frac14}  e^{\frac23 \hbar^{-1} z^{\frac32}} \ e^{\sum_{k=0}^\infty \hbar^k \sum_{2g-2+n=k} \frac{1}{n!}\mathcal F_{g,n}(z,\dots,z) } \cr
&=& z^{-\frac14}  e^{ \frac23 \hbar^{-1} z^{\frac32}} \ e^{\sum_{k=0}^\infty \hbar^k \sum_{2g-2+n=k} \frac{1}{n!} \sum_{d_1+\dots+d_n=3g-3+n} \prod_{i=1}^n \xi_{d_i}(z) } 
\eea
satifies the Airy equation
\beq
\hbar^2 \psi''(x) = x \psi(x).
\eeq

\subsubsection{Summary Painlev\'e 1}


This is the case
\beq
\left\{
\begin{array}{l}
x=z^2-2u  \cr
y = z^3-3uz \cr
y^2-2u^3 = x^3 -3u^2 x
\end{array}
\right.
\eeq
i.e. 
\beq
\td t_1 = -2u \quad , \ \td t_3=6u \quad , \  \td t_5 = -2.
\eeq
\beq
t_5=\td t_5=-2
\quad , \quad
t_3=\td t_3 - 3/2 \td t_5 \td t_1 = 0
\quad , \quad
t_1=\frac{-1}{2}\td t_3\td t_1 + \frac{3}{8}\td t_5 \td t_1^2 = 3u^2 .
\eeq
Painlev\'e 1 is also the minimal model $(3,2)$.

This yields
\beq
T_k =  (3u)^{-k}.
\eeq
and thus
\bea
A[0,0,0] &=& \frac{1}{6u} \cr
D[0] &=& \frac{1}{ 144  u^2}  \quad, \quad D[1] = \frac{1}{144 u} \cr
B[d_1,d_2|d_3] 
&=& \frac{1}{12u} \frac{(2d_3+1)!!}{(2d_1+1)!!(2d_2-1)!!} (3u)^{d_1+d_2-d_3-1} \delta_{d_1+d_2\leq d_3+1} \cr
C[d_1| d_2,d_3] 
&=&  \frac{1}{12u} \frac{(2d_2+1)!!(2d_3+1)!!}{ (2d_1+1)!!} (3u)^{d_1-2-d_2-d_3} \delta_{d_1\leq d_2+d_3+2} .
\eea
The coefficients $F_{g,n}$ are related to the Painlevé 1 equation as follows.
Define
\bea
U 
&=& u - \sum_{g=1}^\infty \hbar^{2g} F_{g,2}[0,0] \cr
&=& u - \frac{\hbar^2}{432 u^4} - \frac{49\hbar^4}{373248 u^9} - \frac{25.49\hbar^6}{2^{11} 3^{9} u^{14}} + O(\hbar^8) \cr
&=& \frac{1}{\sqrt{3}} t_1^{\frac12} - \frac{\hbar^2}{48 } t_1^{-2} - \frac{49\hbar^4}{512.3^{\frac32}} t_1^{-\frac92} - \frac{25.49\hbar^6}{2^{11} 3^{2}} t^{-7} + O(\hbar^8) .
\eea
Regarded as a function of $t_1 = 3u^2$, 
$U$ satisfy the Painlev\'e 1 equation
\bea
 3 U^2 - \frac12 \hbar^2 U''  = t_1.
\eea
In fact this is general: for every minimal model $(p,2)$, $U=-\frac12 \td t_1 - \sum_{g=1}^\infty \hbar^{2g} F_{g,2}[0,0]$ satisfies a non-linear differential equation of the Painlev\'e 1 hierarchy (i.e. a Gelfand-Dikii equation) of order $p-1$ and degree $(p+1)/2$.

\subsubsection{Summary Weil-Petersson}


This is the case where $y$ is the sine function
\beq
\left\{
\begin{array}{l}
x=z^2  \cr
y=\frac{-1}{4\pi} \sin{2\pi z} \cr
\end{array}
\right.
\eeq
i.e. the KdV times
\beq
\td t_{2k+3} = t_{2k+3} =   \frac{(-1)^k (2\pi)^{2k}}{(2k+1)!}
\eeq
which correspond to the $T_k$ being Bernoulli numbers or equivalently even zeta values:
\beq
T_0=1
\quad , \quad
T_k =  \frac{2(2^{2k-1}-1)|B_{2k}|}{(2k)!} (2\pi)^{2k} = 4(2^{2k-1}-1) \zeta(2k) .
\eeq
The $A,B,C,D$ tensors are thus
\bea
A[0,0,0] &=& 1 \cr
D[1] = \frac{1}{24}
\quad , \quad
D[0] &=& \frac{T_1}{8} = \frac12 \zeta(2) = \frac{\pi^2}{12}
 \cr
B[d_1,d_2|d_3] 
&=& \frac{(2d_3+1)!!}{2 (2d_1+1)!!(2d_2-1)!!} T_{d_3+1-d_1-d_2} \cr
C[d_1| d_2,d_3] 
&=& \frac{(2d_2+1)!!(2d_3+1)!!}{2 (2d_1+1)!!} T_{d_2+d_3+2-d_1} .
\eea
The coefficients $F_{g,n}$ are related to Weil-Petersson volumes:
\bea
 F_{g,n}[d_1,\dots,d_n] 
&=& \left<e^{2\pi^2\kappa_1} \prod_{i=1}^n \tau_{d_i}\right>_g \cr
 \sum_{d_1+\dots+d_n \leq 3g-3+n} F_{g,n}[d_1,\dots,d_n] \prod_{i=1}^n \frac{L_i^{2d_i}}{2^{d_i} d_i!} 
 &=& \int_{\modsp_{g,n}(L_1,\dots,L_n)} \!\!\!\! \text{WeilPetersson}  \cr
 &=& \int_{\overline{\modsp}_{g,n}} e^{2\pi^2\kappa_1+\frac12 \sum_{i=1}^n L_i^2 \psi_i}  \cr
 &=& \left<e^{2\pi^2\kappa_1} e^{\frac12 \sum_{i=1}^n L_i^2 \psi_i} \right>_g  .
\eea


\section{GUE - Random matrices}

The following spectral curve appears in random matrices, and in particular applications to enumeration of maps, see for example \cite{EynBook}:

\subsection{Spectral curve}

\beq
\left\{
\begin{array}{l}
\curve = \CC P^1 \cr
x=z+1/z \cr
y = \sum_j u_j z^{-j} \cr
B = \frac{dz_1 dz_2}{(z_1-z_2)^2} \cr
R = \{1,-1\}, \ \sigma_a(z)=1/z.
\end{array}
\right.
\eeq
There are 2 branchpoints $a=\pm 1$, and the involution is $\sigma_a(z)=1/z$.

The spectral curve satisfies an  algebraic equation
\bea
P(x,y) = \left(2y - \sum_j u_j T_j(x) \right)^2 - (x^2-4) \left( \sum_j u_j U_j(x) \right)^2 = 0
\eea
where $T_j(z+z^{-1})=z^j+z^{-j}$ and $U_j(z+z^{-1}) = \frac{z^j-z^{-j}}{z-z^{-1}}$ are Chebyshev polynomials.
$V'(x) = \sum_j u_j T_j(x)$ is the derivative of $V(x)$ called the potential.

The equation is hyperelliptical because of degree $2$ in $y$, but in fact the spectral curve has genus 0 (indeed it is parametrized by rational functions on $\curve = \CC P^1$).

The Gaussian matrix model (GUE) is the case $u_j=\delta_{j,1}$, i.e. $V'(x)=x$, i.e. $V(x) = \frac12 x^2$, in which case the spectral curve obeys the equation
\beq
(2y-x)^2 = (x^2-4)
\eeq
i.e.
\beq
y^2-xy+1=0.
\eeq
$\Im y = \frac12 \sqrt{4-x^2}$ is the equation of a semi-circle of radius $2$, which is the famous Wigner semi-circle spectral curve of the Gaussian unitary matrix model.

The relationship between the Topological Recursion invariants of this spectral curve and a random matrix model is that the expectation value of the trace of the resolvent of a random matrix drawn with probability $e^{-\hbar^{-1}\Tr V(M)}\mathcal D M $ (with $\mathcal D M $ the Lebesgue measure on $H_N$) has the following asymptotic expansion in the limit $\hbar\to 0$, $N\to \infty$ such that $\hbar N = O(1)$: 
\bea
\frac{\int_{H_N} e^{-\hbar^{-1}\Tr V(M)} \mathcal D M \ \   \Tr (x_1-M)^{-1}}{\int_{H_N} e^{-\hbar^{-1}\Tr V(M)} \mathcal D M} \ dx_1 
&=& \sum_{g=0}^\infty \hbar^{2g-2+1} \omega_{g,1}(z_1) ,
\eea
and where $x_1=x(z_1) = z_1+1/z_1$.
A similar relation also holds for any $n>1$, but instead of expectation value, the left hand side is the cumulant expectation value (generalization of covariance). See for example \cite{EynBook} for further details.

\subsubsection{Coordinates}

Instead of $x$ or $z$, it is convenient to use other coordinates:
\beq
x=z+1/z
\quad , \quad
z=e^{2\phi}=\frac{1+\zeta}{1-\zeta}
\quad , \quad
\zeta =\tanh\phi=\sqrt{\frac{x-2}{x+2}}=\frac{z-1}{z+1}
\eeq

\beq
x=z+1/z =2\cosh{2\phi} = 2\frac{1+\zeta^2}{1-\zeta^2} =  \frac{4}{1-\zeta^2}-2
\quad , \quad
dx = \frac{8\zeta d\zeta}{(1-\zeta^2)^2}
\eeq
%
We define the basis of 1-forms with poles at $z=\pm 1$, i.e. at $\zeta = 0,\infty$, for $a=\pm 1$ and $k\in \ZZ_+$:
\bea
d\xi_{a,k}(z) &=& -a(2k+1) \zeta^{-a(2k+1)} \frac{d\zeta}{\zeta} \cr
\xi_{a,k}(z) &=& \zeta^{-a(2k+1)} = \left( \frac{z+1}{z-1} \right)^{a(2k+1)} = \left( \frac{z+a}{z-a} \right)^{2k+1} .
\eea

\subsection{Bergman kernel}

All the following expressions are equivallent for $B$:
\bea
B(z_1,z_2) 
&=& \frac{dz_1 dz_2}{(z_1-z_2)^2} = \frac{d\zeta_1 d\zeta_2}{(\zeta_1-\zeta_2)^2} = \frac{d(1/\zeta_1) d(1/\zeta_2)}{(1/\zeta_1-1/\zeta_2)^2} \cr
&=& d_1\otimes d_2 \ln{(z_1-z_2)} \cr 
&=& d_1\otimes d_2 \ln{(\zeta_1-\zeta_2)} \cr
&=& d_1\otimes d_2 \ln{(1/\zeta_1-1/\zeta_2)} \cr
&=& d_1\otimes d_2 \ln{(1-\zeta_1\zeta_2)} \cr
&=& d_1\otimes d_2 \ln{(1-\zeta_2\zeta_1)} .
\eea

Expansion near $z_2\to a = \pm 1$, i.e. $\zeta_2^a\to 0$
\bea
B(z_1,z_2) 
&=&  d_1\otimes d_2 \ln{(1-\zeta_2^a/\zeta_1^a)}   \cr
&=& - d_1\otimes d_2  \sum_{k=1}^\infty \frac{1}{k} \frac{\zeta_2^{ak}}{\zeta_1^{ak}} \cr
B_{\text{odd}}(z_1,z_2) 
&=& - a  \sum_{k=0}^\infty  \zeta_2^{a(2k+1)-1} d\zeta_2 \ d\xi_{a,k}(z_1) \cr
&=& -   \sum_{k=0}^\infty  \zeta_2^{2ak} d\zeta_2^a \ d\xi_{a,k}(z_1) .
\eea
Integrating over $z_2$ gives
\bea
dS(z_1,z_2) 
&=& - d_1  \sum_{k=1}^\infty \frac{1}{k} \frac{\zeta_2^{ak}}{\zeta_1^{ak}} \cr
dS_{\text{odd}}(z_1,z_2) 
&=& - d_1  \sum_{k=0}^\infty \frac{1}{2k+1} \frac{\zeta_2^{a(2k+1)}}{\zeta_1^{a(2k+1)}} \cr
&=& - \sum_{k=0}^\infty \frac{1}{2k+1} \zeta_2^{a(2k+1)} \ d\xi_{a,k}(z_1)  .
\eea

\subsubsection{Kernel}

The kernel is
\beq
K_a(z_1,z) = \frac{dS_{\text{odd}}(z_1,z)}{(y(z)-y(1/z))dx(z)}.
\eeq
The denominator is
\bea
&& (y(z)-y(1/z))dx(z) \cr
&=& -\sum_{j\geq 1} u_j (z^j-z^{-j})dx(z) \cr
&=& - \frac{8  \zeta d\zeta}{(1-\zeta^2)^2} \sum_{j\geq 1} u_j a^j ((1+\zeta^a)^j(1-\zeta^a)^{-j} - (1-\zeta^a)^j(1+\zeta^a)^{-j} ) \cr
&=& -8  \zeta d\zeta\sum_{j\geq 1} u_j a^j \left( \frac{(1+\zeta^a)^{2j}}{(1-\zeta^{2a})^j}- \frac{(1-\zeta^a)^{2j}}{(1-\zeta^{2a})^j} \right)(1-\zeta^2)^{-2} \cr
&=& -8  \zeta d\zeta\sum_{j\geq 1} u_j a^j \left( (1+\zeta^a)^{2j}-(1-\zeta^a)^{2j} \right)(1-\zeta^{2a})^{-2-j} \frac{(1-\zeta^{2a})^2}{(1-\zeta^2)^2} \cr
&=& -8  \zeta d\zeta\sum_{j\geq 1} u_j a^j \left( (1+\zeta^a)^{2j}-(1-\zeta^a)^{2j} \right)(1-\zeta^{2a})^{-2-j} (a \zeta^{a-1})^2 \cr
&=& -16  \zeta^{2a} \frac{d\zeta}{\zeta}\sum_{j\geq 1} u_j a^j \left(\sum_{l=0}^{j-1} \begin{pmatrix} 2j \cr 2l+1 \end{pmatrix} \zeta^{a(2l+1)} \right)(1-\zeta^{2a})^{-2-j} \cr
&=& -16  \zeta^{3a} \frac{d\zeta}{\zeta}\sum_{j\geq 1} u_j a^j \left(\sum_{l=0}^{j-1} \begin{pmatrix} 2j \cr 2l+1 \end{pmatrix} \zeta^{2al} \right)  \left( \sum_{k\geq 0} \begin{pmatrix} -j-2 \cr k \end{pmatrix} (-1)^k \zeta^{2ak} \right) \cr
&=& -16 a \zeta^{3a} \frac{d\zeta^a}{\zeta^a}\sum_{j\geq 1} u_j a^j \left(\sum_{l=0}^{j-1} \begin{pmatrix} 2j \cr 2l+1 \end{pmatrix} \zeta^{2al} \right)  \left( \sum_{k\geq 0} \begin{pmatrix} -j-2 \cr k \end{pmatrix} (-1)^k \zeta^{2ak} \right) \cr
&=& -16  a\zeta^{3a} \frac{d\zeta^a}{\zeta^a} 
\sum_{k\geq 0} \zeta^{2ak} 
\sum_{j\geq 1} u_j a^j \left(\sum_{l=0}^{j-1} \begin{pmatrix} 2j \cr 2l+1 \end{pmatrix}   \begin{pmatrix} -j-2 \cr k-l \end{pmatrix} (-1)^{k-l}  \right) \cr
&=& -16  a \zeta^{3a} \frac{d\zeta^a}{\zeta^a} 
\sum_{k\geq 0} \zeta^{2ak} 
\sum_{j\geq 1} u_j a^j \left(\sum_{l=0}^{j-1} \begin{pmatrix} 2j \cr 2l+1 \end{pmatrix}   \begin{pmatrix} j+1+k-l \cr j+1 \end{pmatrix}   \right) \cr
\eea
Let us write it
\bea
 (y(z)-y(1/z))dx(z) 
&=& -16 a  \zeta^{2a} d\zeta^a 
\sum_{k\geq 0} \zeta^{2ak} y_{a,k} \cr
y_{a,k} &=& \sum_{j\geq 1} u_j a^j \left(\sum_{l=0}^{\min(j-1,k)} \begin{pmatrix} 2j \cr 2l+1 \end{pmatrix}   \begin{pmatrix} j+1+k-l \cr j+1 \end{pmatrix}   \right) \cr
\eea
and let's write its inverse as
\bea
\frac{1}{ (y(z)-y(1/z))dx(z) }
&=& - \frac{a}{16  \zeta^{2a} d\zeta^a }
\frac{1}{y_{a,0}} \left( 1+ \sum_{k\geq 1} \zeta^{2ak} y_{a,k}/y_{a,0} \right)^{-1} \cr
&=& - \frac{a}{16   \zeta^{2a} d\zeta^a } \left( \sum_{k\geq 0} \zeta^{2ak} Y_{a,k} \right) 
\eea
where
\bea
Y_{a,0} &=& 1/y_{a,0} \cr
Y_{a,k} &=& \sum_{n= 1}^k \frac{(-1)^n}{y_{a,0}^{n+1}} \sum_{j_1+\dots+j_n=k, \ j_i>0} y_{a,j_1}\dots y_{a,j_n}  \qquad k\geq 1 .
\eea
Therefore
\bea 
K_a(z_1,z)
&=&  \frac{a}{16   \zeta^{2a} d\zeta^a } \left(  \sum_{k\geq 0} \zeta^{2ak} Y_{a,k} \right) 
\left( \sum_{d_1=0}^\infty \frac{1}{2d_1+1} \zeta^{a(2d_1+1)} \ d\xi_{a,d_1}(z_1) \right) \cr
&=&  \frac{a}{16   \zeta^{a} d\zeta^a } \left( \sum_{k=0} \zeta^{2ak} \sum_{j=0}^k  Y_{a,k-j} \frac{d\xi_{a,j}(z_1)}{2j+1}   \right) .
\eea

\subsection{ABCD Tensors}

From the definitions we get

$\bullet$ $A$
\bea
\omega_{0,3}(z_1,z_2,z_3)
&=& \sum_{a=\pm 1} 2\Res_{a} K_a(z_1,z) B(z,z_2) B(1/z,z_3)  \cr
&=& \sum_{a=\pm 1} \Res_{a} \frac{a}{8 y_{a,0}  \zeta^{a} d\zeta^a } \left( d\xi_{a,0}(z_1) + O(\zeta^{2a})    \right) \cr 
&& \qquad \left( - d\zeta^a d\xi_{a,0}(z_2) \right) \left(  d\zeta^a d\xi_{a,0}(z_3) \right) \cr
&=& - \sum_{a=\pm 1}  \frac{a}{8y_{a,0}}  d\xi_{a,0}(z_1)  d\xi_{a,0}(z_2)  d\xi_{a,0}(z_3)  \cr
\eea
i.e.
\beq
A[a,0;a,0;a,0] = F_{0,3}[a,0;a,0;a,0] = \frac{-a Y_{a,0}}{8} =\frac{-a}{8 y_{a,0}}
\eeq

$\bullet$ $D$
\bea
\omega_{1,1}(z_1)
&=& \sum_{a=\pm 1} \Res_{a} K_a(z_1,z) B(z,1/z)   \cr
&=& \sum_{a=\pm 1} \Res_{0}  \frac{a}{16   \zeta^{a} d\zeta^a } ( Y_{a,0} + \zeta^{2a}Y_{a,1}) (d\xi_{a,0}(z_1) + \frac13\zeta^{2a} d\xi_{a,1}(z_1) )     \frac{- (d\zeta^a)^2}{4 \zeta^{2a}}   \cr
&=& - \sum_{a=\pm 1}  \frac{a}{64 }   \left( Y_{a,1}d\xi_{a,0}(z_1) + \frac{Y_{a,0}}3  d\xi_{a,1}(z_1) \right)      \cr
\eea
\beq
D[a,0] = \frac{-a Y_{a,1}}{64}
\quad , \quad
D[a,1] = \frac{-a Y_{a,0}}{192 }
\eeq

$\bullet$ $C$
\bea
&& \Res_{a_1} K_{a_1}(z_1,z) d\xi_{a_2,d_2}(z) d\xi_{a_3,d_3}(1/z)  \cr
&=&  - \Res_{0}  \frac{a_1}{16 \zeta^{a_1} d\zeta^{a_1} } \left( \sum_k \zeta^{2a_1 k} Y_{a_1,k} \right) \cr 
&& \qquad \left( \sum_{d_1} \frac{\zeta^{2a_1 d_1}}{2d_1+1}  d\xi_{a_1,d_1}(z_1) \right) 
d\xi_{a_2,d_2}(z) d\xi_{a_3,d_3}(z) \cr
\eea
\bea
&& C[a_1,d_1 | a_2,d_2; a_3,d_3] \cr
&=& - \Res_{0}  \frac{a_1}{16  \zeta^{a_1} d\zeta^{a_1} } \left( \sum_k \zeta^{2a_1 k} Y_{a_1,k} \right)  \frac{\zeta^{2a_1 d_1}}{2d_1+1}  
d\xi_{a_2,d_2}(z) d\xi_{a_3,d_3}(z)  \cr
&=& - a_1 a_2 a_3 \frac{(2d_2+1)(2d_3+1)}{2d_1+1} \Res_{0}  \frac{1}{16  \zeta^{a_1} d\zeta^{a_1} } \left( \sum_k \zeta^{2a_1 k} Y_{a_1,k} \right) \cr  && \qquad \zeta^{2a_1 d_1} \zeta^{-a_2(2d_2+1)} \zeta^{-a_3(2d_3+1)} \frac{(d\zeta^{a_1})^2}{\zeta^{2a_1}}  \cr
&=& - a_1 a_2 a_3 \frac{(2d_2+1)(2d_3+1)}{2d_1+1} \Res_{0}  \frac{d \zeta^{a_1}}{16   \zeta^{a_1}  } \left( \sum_k \zeta^{2a_1 k} Y_{a_1,k} \right)  \cr 
&& \qquad \zeta^{2a_1 (d_1-1)} \zeta^{-a_2(2d_2+1)} \zeta^{-a_3(2d_3+1)}   \cr
&=& \frac{-1}{16 } a_1 a_2 a_3 \frac{(2d_2+1)(2d_3+1)}{2d_1+1}   Y_{a_1,a_1 a_2 d_2 + a_1 a_3 d_3 - d_1 +1 + a_1(a_2+a_3)/2}  \cr
\eea
it is non-zero only if
\beq
d_1 \leq 1+ a_1 (a_2 d_2 +  a_3 d_3   + (a_2+a_3)/2 )
\eeq

$\bullet$ $B$
\bea
&& B[a,d_1;a,d_2 | a_3 d_3] \cr
&=& -\Res_{0} \frac{a}{16  \zeta^a d\zeta^a} \left(\sum_k \zeta^{2ak} Y_{a,k} \right) \frac{\zeta^{2ad_1}}{2d_1+1} (- \zeta^{2ad_2} d\zeta^a) d\xi_{a_3,d_3}(z) \cr
&=& \frac{a}{16 (2d_1+1)} \Res_{0} \frac{1}{\zeta^a d\zeta^a} \left(\sum_k \zeta^{2ak} Y_{a,k} \right) \zeta^{2ad_1} \zeta^{2ad_2} d\zeta^a d\xi_{a_3,d_3}(z) \cr
&=& \frac{a}{16 (2d_1+1)} \Res_{0} \frac{1}{\zeta^a } \left(\sum_k \zeta^{2ak} Y_{a,k} \right) \zeta^{2ad_1} \zeta^{2ad_2} \zeta^{-a_3(2d_3+1)} (-a_3(2d_3+1)) \frac{d\zeta}{\zeta} \cr
&=& \frac{1}{16 (2d_1+1)} \Res_{0} \frac{1}{\zeta^a } \left(\sum_k \zeta^{2ak} Y_{a,k} \right) \zeta^{2ad_1} \zeta^{2ad_2} \zeta^{-a_3(2d_3+1)} (-a_3(2d_3+1)) \frac{d\zeta^a}{\zeta^a} \cr
&=& \frac{-a_3(2d_3+1)}{16 (2d_1+1)} \Res_{0}  \left(\sum_k \zeta^{2ak} Y_{a,k} \right) \zeta^{2a(d_1+d_2-a a_3 d_3)}  \zeta^{-a(1+a a_3)}  \frac{d\zeta^a}{\zeta^a} \cr
&=& \frac{-a_3(2d_3+1)}{16 (2d_1+1)}   Y_{a,a a_3 d_3-d_1-d_2 +(1+a a_3)/2}  \cr
\eea

\subsection{Summary GUE}

\beq
\left\{
\begin{array}{l}
x=z+1/z \cr
y = \sum_j u_j z^{-j} \cr
B = \frac{dz_1 dz_2}{(z_1-z_2)^2} \cr
a=\pm 1
\end{array}
\right.
\eeq
We define for the ramification points $a=\pm 1$:
\bea
y_{a,k} &=& \sum_{j\geq 1} u_j a^j \left(\sum_{l=0}^{\min(j-1,k)} \begin{pmatrix} 2j \cr 2l+1 \end{pmatrix}   \begin{pmatrix} j+1+k-l \cr j+1 \end{pmatrix}   \right) \cr
Y_{a,k} &=& \sum_{n= 0}^k \frac{(-1)^n}{y_{a,0}^{n+1}} \sum_{j_1+\dots+j_n=k, \ j_i>0} y_{a,j_1}\dots y_{a,j_n}  \qquad k\geq 1 \cr
\eea
In particular
\bea
y_{a,0} &=& 2 \sum_{j\geq 1} j u_j a^j 
\quad , \quad
Y_{a,0} = \frac{1}{y_{a,0}} .
\eea

\bea
A[a,0;a,0;a,0] &=& \frac{-a}{8} Y_{a,0} \cr
D[a,1] &=& \frac{-a}{192} Y_{a,0} \cr
D[a,0] &=& \frac{-a}{64} Y_{a,1} \cr
B[a,d_1;a,d_2|a_3,d_3] &=& \frac{-a_3(2d_3+1)}{16 (2d_1+1)}   Y_{a,a a_3 d_3-d_1-d_2 +(1+a a_3)/2}  \cr
C[a_1,d_1|a_2,d_2;a_3,d_3] &=& \frac{-a_1 a_2 a_3}{16 }  \frac{(2d_2+1)(2d_3+1)}{2d_1+1}  \cr
&& \qquad Y_{a_1,a_1 a_2 d_2 + a_1 a_3 d_3 - d_1 +1 + a_1(a_2+a_3)/2}  \cr
\eea

For pure GUE ($u_j=\delta_{j,1}$) we have
\bea
y_{a,k} &=& (k+2)(k+1)a \cr
Y_{a,k} &=& a \frac{(-1)^k}{2} \binomial{3}{k}
\eea

\bea
A[a,0;a,0;a,0] &=& \frac{-1}{16}  \cr
D[a,1] &=& \frac{-1}{384}  \cr
D[a,0] &=& \frac{3}{128}  \cr
B[a,d_1;a,d_2|a_3,d_3] &=& \frac{-aa_3(2d_3+1)}{32 (2d_1+1)} (-1)^k  \binomial{3}{k} \cr 
&& \qquad \quad \text{with } \ k=a a_3 d_3-d_1-d_2 +(1+a a_3)/2  \cr
C[a_1,d_1|a_2,d_2;a_3,d_3] &=& \frac{- a_2 a_3}{32 }  \frac{(2d_2+1)(2d_3+1)}{2d_1+1} (-1)^k \binomial{3}{k} \cr
&& \  \text{with }  \  k=a_1 a_2 d_2 + a_1 a_3 d_3 - d_1 +1 + a_1(a_2+a_3)/2  . \cr
\eea











\section{Elliptic curves}

Consider $\tau\in \CC$ such that $\Im\tau>0$.
Let $\curve$ be the torus of modulus $\tau$:
\beq
\curve = \CC/\ZZ+\tau\ZZ . 
\eeq
An elliptic function $f$ is such that $f(z)=f(z+1)=f(z+\tau)$.
We recall that the only elliptic holomorphic functions are constant. 
Non-constant elliptic meromorphic functions must have at least 2 simple poles or at least one higher degree pole. 
The sum of residues at all poles must be 0.

\subsection{Elliptic spectral curves}

We shall consider the following class of spectral curves:
\beq
\left\{
\begin{array}{l}
\curve = \CC/\ZZ+\tau\ZZ \cr
z\mapsto x(z) = \text{elliptic meromorphic function} \cr
z\mapsto y(z) = \text{elliptic meromorphic function} \cr
B = (\wp(z_1-z_2)+G_2)dz_1\otimes dz_2 \cr
 \sigma_a(z)=2a-z
\end{array}
\right.
\eeq
where $\wp$ is the Weierstrass elliptic function (see below), $G_2\in \CC$ is arbitrary (often one chooses $G_2=G_2(\tau)$ the 2nd Eisenstein series (in which case $B$ is normalized on the $\acycle$-cycle = $\RR / \ZZ+\tau\ZZ$ ), but any other choice is good too).
We also assume that the ramification points are simple, and that the involutions take the form:
\beq
\sigma_a(z) = 2a-z.
\eeq

\subsubsection{Reminder Weierstrass elliptic function}

Recall that the Weierstrass function is:
\bea
\wp(z) 
&=& \frac{1}{z^2} + \sum_{(n,m)\in \ZZ^2\setminus \{(0,0)\}} \frac{1}{(n+\tau m-z)^{2}}-\frac{1}{(n+\tau m)^{2}} \cr
&=& \frac{1}{z^2} + \sum_{k\geq 2} (2k-1)G_{2k} z^{2k-2} \cr
&& \text{where } G_{2k} = \sum_{(n,m)\in \ZZ^2\setminus \{(0,0)\}} \frac{1}{(n+\tau m)^{2k}}  = 2k\text{-th Eisenstein series}.
\eea
It satisfies the differential equation
\bea
\wp'^2 &=&   4(\wp^3 - 15 G_4 \wp - 35 G_6) ,
\eea
from which we deduce
\bea
\wp'' &=& 6 (\wp^2 - 5 G_4) \cr
\wp''' &=& 12\wp \wp' .
\eea
And in general, even derivatives $\wp^{(2k)}$ can be expressed as polynomials of $\wp$:
\bea\label{eq:wpderiveQ}
\wp^{(2k)} &=& (2k+1)! Q_k(\wp) \cr
\eea
where $Q_k$ is a polynomial of degree $k+1$.
It satisfies the recursion
\beq
Q_0(x) = x \quad , \quad
(2k+3)(2k+2) Q_{k+1}(x) = 6(x^2-5G_4)Q'_k + 4(x^3-15G_4 x-35 G_6)Q''_k .
\eeq
It can be computed with the following formula
\bea\label{eq:computewpderivQ}
Q_k(x) &=&  \left( x^{k+1} + \sum_{j=0}^{k-1} \alpha_{k,j} x^{j} \right) \cr
\alpha_{k,j} &=& \frac{k+1}{j} [z^{2(k+1-j)}] \left(1+\sum_{l=2}^{k-j} (2l-1) G_{2l}z^{2l} \right)^{-j} \cr
\alpha_{k,0} &=& G_{2k+2} -(k+1) [z^{2k+2}] \ln\left(1+\sum_{l=2}^{k} (2l-1) G_{2l}z^{2l} \right) \cr
Q_{k}(x) &=& \left( G_{2k+2} - (k+1) [z^{2k+2}] \ln{(z^2 \wp(z)-z^2 x)} \right) . 
\eea

Given an arbitrary $G_2\in\CC$, it is convenient to define:
\beq
\hat\wp(z) =\wp(z) + G_2.
\eeq

\subsubsection{Reminder Legendre-Jacobi elliptic functions}

The Legendre normalization of an elliptic curve is the algebraic equation
$$ 
y^2 = (1-x^2)(1-k^2 x^2) .
$$
Then, define the Legendre elliptic integrals
\bea
K &=& K(k) = \int_0^1 \frac{dx}{\sqrt{(1-x^2)(1-k^2 x^2)}} \cr
K' &=& K(\sqrt{1-k^2}) \cr
\tau &=& \frac{i K'}{K} \cr
F(x) &=& \int_0^x \frac{dx'}{\sqrt{(1-x'^2)(1-k^2 x'^2)}} .
\eea
The Jacobi elliptic functions are then defined as
\bea
\sn(u) &=& F^{-1}(u) \cr
\cn(u) &=& \sqrt{1-\sn(u)^2} \cr
\dn(u) &=& \sqrt{1-k^2 \sn(u)^2} 
\eea
They satisfy many nice properties, that generalize those of trigonometric functions, $K$ is like $\pi/2$, and:
\bea
\sn(-u) &=& -\sn(u) \cr
\sn(u+4K) &=& \sn(u) \cr
\sn(u+2K) &=& -\sn(u) \cr
\sn(u+2iK') &=& \sn(u) \cr
&& \cr
\sn(K) &=&  1 =-\sn(3K) \cr
\sn(2K) &=&  0 \cr
\sn(iK') &=&  \infty \cr
&& \cr
\cn(-u) &=& \cn(u) \cr
\cn(u+4K) &=& \cn(u) \cr
\cn(u+2K) &=& -\cn(u) \cr
\cn(u+2iK') &=& -\cn(u) \cr
&& \cr
\cn(0) &=&  1 \cr
\cn(K) &=&  0 =\cn(3K) \cr
\cn(2K) &=&  -1 \cr
\cn(iK') &=&  \infty \cr
&& \cr
\dn(-u) &=& \dn(u) \cr
\dn(u+2K) &=& \dn(u) \cr
\dn(u+4iK') &=& \dn(u) \cr
\dn(u+2iK') &=& -\dn(u) \cr
&& \cr
\dn(0) &=&  1 \cr
\dn(K) &=&  \sqrt{1-k^2}  \cr
\dn(iK') &=&  \infty \cr
\eea

\bea
\sn'(u) &=& \cn u \dn u \cr
\cn'(u) &=& -\sn u \dn u \cr
\dn'(u) &=& -k^2 \sn u \cn u .
\eea

We have the relationship to the Weierstrass function
\bea
\sn(2K z)^2 &=& \frac{\wp(1/2)-\wp(\tau/2)}{\wp(z)-\wp(\tau/2)} \cr
\text{where } & & k^2 = \frac{\wp(1/2+\tau/2)-\wp(\tau/2)}{\wp(1/2)-\wp(\tau/2)} .
\eea

\subsection{Basis}

We choose the basis:
\beq
a\in R, \ d\in \ZZ_+ \ : \quad
d\xi_{a,d}(z) = (\wp^{(2d)}(z-a)+G_2\delta_{d,0}) dz = \hat\wp^{(2d)}(z-a) dz .
\eeq
It is odd under involutions $z\mapsto \sigma_a(z)=2a-z$.

It has the expansion near $a$:
\bea
d\xi_{a,d}(z) 
&\sim_{z\to a}& \frac{(2d+1)!}{(z-a)^{2d+2}} + \sum_{j=0}^\infty   G_{2d+2j+2}\frac{(2d+2j+1)!}{(2j)!} \ (z-a)^{2j} \cr
&\sim_{z\to a}& \frac{(2d+1)!}{(z-a)^{2d+2}} \left( 1 + \sum_{j=0}^\infty   G_{2d+2j+2} \binomial{2d+2j+1}{2j} \ (z-a)^{2d+2j+2} \right) \cr
\eea
It has the expansion near $b\neq a$:
\bea
d\xi_{a,d}(z) 
&\sim_{z\to b}& \sum_{j=0}^\infty \frac{1}{(2j)!} \hat\wp^{(2d+2j)}(b-a)\ (z-b)^{2j} \cr
&\sim_{z\to b}& \sum_{j=0}^\infty \frac{1}{(2j)!} ((2d+2j+1)! Q_{d+j}(\wp(b-a)) + \delta_{d+j,0}G_2) \ (z-b)^{2j} . \cr
\eea
The Kernel $B$ has expansion
\bea
B(z_1,z)
&\sim_{z\to a}& \sum_k \frac{(z-a)^{k}}{k!} \hat\wp^{(k)}(a-z_1) dz dz_1 \cr
B_{\text{odd}}(z_1,z)
&\sim_{z\to a}& \sum_d \frac{(z-a)^{2d}dz}{(2d)!} d\xi_{a,d}(z_1) \cr
dS_{\text{odd}}(z_1;z)
&\sim_{z\to a}& \sum_d \frac{(z-a)^{2d+1}}{(2d+1)!} d\xi_{a,d}(z_1) \cr
\eea

\subsubsection{Kernel}


\bea
K_a(z_1,z)
&=& \frac{1}{(z-a)dz} \left(\sum_{k\geq 0} Y_{a,k}  (z-a)^{2k} \right) \sum_{d\geq 0} \frac{d\xi_{a,d}(z_1)}{(2d+1)!} (z-a)^{2d}  \cr
\eea
where
\bea
\frac{1}{(y(z)-y(\sigma_a(z))) x'(z)}
&=& \sum_{k=0}^\infty Y_{k,a} (z-a)^{2k-2}.
\eea
In particular
\bea
Y_{a,0} = \frac{1}{2 y'(a) x''(a)}
\eea

$\bullet$ $A$
\bea
\omega_{0,3}(z_1,z_2,z_3)
&=& -2\sum_{a\in R}  \Res_a K_a(z_1,z) B(z,z_2)B(z,z_3) \cr
&=& -2\sum_{a\in R}  \frac{1}{(z-a)dz} \left(Y_{a,0} + O((z-a)^2) \right) \left( d\xi_{a,0}(z_1) + O((z-a)^2)\right)  \cr 
&& \left( d\xi_{a,0}(z_2) + O((z-a)^2)\right) \left( d\xi_{a,0}(z_3) + O((z-a)^2)\right) dz^2 \cr
&=& -2\sum_{a\in R}  Y_{a,0} \prod_{i=1}^3 d\xi_{a,0}(z_i)
\eea
\bea
A[a,0;a,0;a,0] &=& -2 Y_{a,0} \cr
\eea

$\bullet$ $D$
\bea
B(z,\sigma_a(z))
&=& - (\wp(2z)+G_2) dz^2 \cr
&\sim_{z\to a}& - \left( \frac{1}{4(z-a)^2}+G_2 +O((z-a)^2) \right) dz^2 \cr
\eea

\bea
\omega_{1,1}(z_1)
&=& - \sum_{a\in R}  \frac{1}{(z-a)dz} \left(Y_{a,0} + (z-a)^2 Y_{a,1} + O((z-a)^4) \right) \cr 
&& \left(d\xi_{a,0}(z_1) + \frac16 (z-a)^2 d\xi_{a,1}(z_1) + O((z-a)^4)\right) \cr 
&& \left( \frac{1}{4(z-a)^2}+G_2 +O((z-a)^2) \right) dz^2 \cr
&=& - \frac14 \sum_{a\in R}  \frac{dz}{(z-a)^3 } \left(Y_{a,0} + (z-a)^2 Y_{a,1} + O((z-a)^4) \right) \cr 
&& \left(d\xi_{a,0}(z_1) + \frac16 (z-a)^2 d\xi_{a,1}(z_1) + O((z-a)^4)\right) \cr 
&& \left( 1+ 4 (z-a)^2 G_2 +O((z-a)^4) \right)  \cr
&=& - \frac14 \sum_{a\in R} \frac16 Y_{a,0} d\xi_{a,1}(z_1) + Y_{a,1}  d\xi_{a,0}(z_1) + 4G_2 d\xi_{a,0}(z_1) \cr
\eea
\bea
D[a,1] &=& \frac{-Y_{a,0}}{24} \cr
D[a,0] &=& \frac{-1}{4} (Y_{a,1} + 4 G_2 Y_{a,0})\cr
\eea

$\bullet$ $B$

\bea
&& B[a,d_1;a,d_2|a_3,d_3] \cr
&=& -\Res_a \frac{1}{(z-a)dz} \left(\sum_{k\geq 0} Y_{a,k}  (z-a)^{2k} \right) \frac{(z-a)^{2d_1}}{(2d_1+1)!} \frac{(z-a)^{2d_2}}{(2d_2)!} dz d\xi_{a_3,d_3}(z)  \cr
&=& - \frac{1}{(2d_1+1)! (2d_2)!} \Res_a \left(\sum_{k\geq 0} Y_{a,k}  (z-a)^{2k} \right) (z-a)^{2d_1+2d_2-1} d\xi_{a_3,d_3}(z)  \cr
&=& \frac{-1}{(2d_1+1)! (2d_2)!} \Res_a \left(\sum_{k\geq 0} Y_{a,k}  (z-a)^{2(k+d_1+d_2)-1} \right)  \hat\wp^{(2d_3)}(z-a_3) dz \cr
\eea
If $a_3\neq a$ this gives 0 except if $0=k=d_1=d_2$, and then
\bea
 B[a,0;a,0|a_3,d_3]  
 &=& - \frac{Y_{a,0}}{2}  \hat\wp^{(2d_3)}(a-a_3) \cr
&=& - \frac{Y_{a,0}}{2}  ( \wp^{(2d_3)}(a-a_3)+\delta_{d_3,0}G_2)  \cr
\eea
If $a_3= a$ this gives
\bea
&& B[a,d_1;a,d_2|a,d_3] \cr
&=& \frac{-1}{(2d_1+1)! (2d_2)!} \Res_a \left(\sum_{k\geq 0} Y_{a,k}  (z-a)^{2(k+d_1+d_2)-1} \right)  \hat\wp^{(2d_3)}(z-a)  dz \cr
&=& \frac{-1}{(2d_1+1)! (2d_2)!} \Res_a \left(\sum_{k\geq 0} Y_{a,k}  (z-a)^{2(k+d_1+d_2)-1} \right) dz \cr  
&& \left( \frac{(2d_3+1)!}{(z-a)^{2d_3+2}} + \sum_{j\geq d_3+1} (2j-1)G_{2j} (z-a)^{2j-2-2d_3} \frac{(2j-2)!}{(2j-2-2d_3)!} \right)  \cr
&=& \frac{-(2d_3+1)!}{(2d_1+1)! (2d_2)!} Y_{a,d_3+1-d_1-d_2} 
-(2d_3+1)! G_{2d_3+2} Y_{a,0} \delta_{d_1,0}\delta_{d_2,0} 
\eea

\beq
 B[a,d_1;a,d_2|a,d_3] = - \frac{ (2d_3+1)! }{(2d_1+1)! (2d_2)!}   \left( Y_{a,1+d_3-d_1-d_2} +  G_{2d_3+2} Y_{a,0} \delta_{d_1,0}\delta_{d_2,0} \right) .
\eeq

$\bullet$ $C$

\bea
&& C[a_1,d_1|a_2,d_2;a_3,d_3]  \cr
&=& - \Res_{z\to a_1} \frac{dz}{z-a_1} \left(\sum_{k\geq 0} Y_{a_1,k}  (z-a_1)^{2k} \right)  \frac{(z-a_1)^{2d_1}}{(2d_1+1)!} \hat\wp^{(2d_2)}(z-a_2) \hat\wp^{(2d_3)}(z-a_3)  \cr
&=& \frac{-1}{(2d_1+1)!} \Res_{z\to 0} \left(\sum_{k\geq 0} Y_{a_1,k}  z^{2(k+d_1)-1} \right) dz \hat\wp^{(2d_2)}(z+a_1-a_2) \hat\wp^{(2d_3)}(z+a_1-a_3)  \cr
\eea
If $a_2\neq a_1$ and $a_3\neq a_1$:
\bea
 C[a_1,d_1|a_2,d_2;a_3,d_3]  
&=& - Y_{a_1,0} \delta_{d_1,0} \hat\wp^{(2d_2)}(a_1-a_2) \hat\wp^{(2d_3)}(a_1-a_3) 
\eea
If $a_2= a_1$ and $a_3\neq a_1$:
\bea
&& C[a_1,d_1|a_1,d_2;a_3,d_3]  \cr
&=& \frac{-1}{(2d_1+1)!} \Res_{z\to 0} \left(\sum_{k\geq 0} Y_{a_1,k}  z^{2(k+d_1)-1} \right) dz \hat\wp^{(2d_2)}(z) \hat\wp^{(2d_3)}(z+a_1-a_3)  \cr
&=& \frac{-(2d_2+1)!}{(2d_1+1)!} \Res_{z\to 0} \left(\sum_{k\geq 0} Y_{a_1,k}  z^{2(k+d_1)-1} \right) dz \cr
&& \left(\frac{1}{z^{2d_2+2}} + G_{2d_2+2} + O(z^2) \right) \hat\wp^{(2d_3)}(z+a_1-a_3)  \cr
&=& \frac{-(2d_2+1)!}{(2d_1+1)!} \Big( \delta_{d_1,0}  G_{2d_2+2} Y_{a_1,0} \hat\wp^{(2d_3)}(a_1-a_3) \cr 
&& + \sum_{k=0}^{d_2+1-d_1} \frac{1}{(2k)!} Y_{a_1,d_2+1-d_1-k} \hat\wp^{(2d_3+2k)}(a_1-a_3) 
\Big) \cr
\eea

If $a_2=a_3= a_1$:
\bea
&& C[a_1,d_1|a_1,d_2;a_1,d_3]  \cr
&=& \frac{-1}{(2d_1+1)!} \Res_{z\to 0} \left(\sum_{k\geq 0} Y_{a_1,k}  z^{2(k+d_1)-1} \right) dz \hat\wp^{(2d_2)}(z) \hat\wp^{(2d_3)}(z)  \cr
&=& \frac{-1}{(2d_1+1)!} \Res_{z\to 0} \left(\sum_{k\geq 0} Y_{a_1,k}  z^{2(k+d_1)-1} \right) dz   \cr
&& \left(\frac{(2d_2+1)!}{z^{2d_2+2}} + \sum_{j=0}^{d_3+1} G_{2d_2+2+2j} z^{2j} \frac{(2d_2+2j+1)!}{(2j)!} + O(z^{2d_3+4}) \right) \cr
&& \left(\frac{(2d_3+1)!}{z^{2d_3+2}} + \sum_{j=0}^{d_2+1} G_{2d_3+2+2j} z^{2j} \frac{(2d_3+2j+1)!}{(2j)!} + O(z^{2d_2+4}) \right) \cr
&=& - \frac{(2d_2+1)!(2d_3+1)!}{(2d_1+1)!} \Big(  Y_{a_1,d_2+d_3+2-d_1} \cr 
&& + \sum_{j=0}^{d_3+1-d_1} G_{2d_2+2+2j} \binomial{2d_2+2j+1}{2j} Y_{a_1,d_3+1-d_1-j} \cr
&& + \sum_{j=0}^{d_2+1-d_1} G_{2d_3+2+2j} \binomial{2d_3+2j+1}{2j} Y_{a_1,d_2+1-d_1-j} \cr
&& +\delta_{d_1,0} Y_{a_1,0} G_{2d_2+2} G_{2d_3+2} 
\Big) \cr
\eea

\subsection{Weierstrass elliptic curve}

This is the case 
\beq
\begin{cases}
y^2 = 4(x^3 - 15 G_4 x - 35 G_6) \cr
x(z) = \wp(z) \cr
y(z) = \wp'(z) \cr
R = \left\{ \frac12 , \frac12+\frac{\tau}{2}, \frac{\tau}{2} \right\} , \ \sigma_a(z)=-z.
\end{cases}
\eeq
We have 
\bea
\wp(a) = x_a \ \quad \text{such that } x_a^3 = 15 G_4 x + 35 G_6.
\eea
We have
\bea
\sum_{k=0}^\infty Y_{a,k} \ (z-a)^{2k-2}
&=&  \frac{1}{2\wp'(z)^2} \cr
\sum_{k=0}^\infty Y_{a,k} \ (z-a)^{2k}
&=&  \frac12 \left( \sum_{k=0}^\infty Q_{k+1}(x_a) (z-a)^{2k} \right)^{-2} \cr
\eea

\bea\label{eq:YakforWeierstrass}
Y_{a,0} &=& \frac{1}{2\wp''(a)^2} = \frac{1}{2 Q_1(x_a)^2} = \frac{1}{72(x_a^2-5 G_4)^2}\cr
Y_{a,1} &=&  -\frac{Q_2(x_a)}{Q_1(x_a)^3} \cr
Y_{a,2} &=&  -\frac{Q_3(x_a)}{Q_1(x_a)^3}  + \frac32\frac{Q_2(x_a)^2}{Q_1(x_a)^4} \cr
Y_{a,k} &=&   \sum_{l=1}^k (-1)^l \frac{l+1}{2} Q_1(x_a)^{-l-2}  \sum_{k_1+\dots+k_l=k, \ k_i\geq 1} \ \prod_{i=1}^l Q_{k_i+1}(x_a) .
 \eea

\subsection{Legendre elliptic curve}

This is the case
\beq
\left\{
\begin{array}{l}
y^2 = (1-x^2)(1-k^2 x^2) \cr
x= \sn(2K z)  \cr
y = \cn(2K z)\dn(2K z) \cr
R = \{\frac12,\frac12+\frac12\tau\} 
\end{array}
\right.
\eeq

We have for $a=1/2$ or $a=1/2+\tau/2$:
\bea
\sum_{k=0}^\infty Y_{a,k} \ (z-a)^{2k-2}
&=&  \frac{1}{4K \cn(2K z)^2\dn(2K z)^2} \cr
&=&  \frac{1}{4K} \frac{\left( \wp(z)-\wp(\tau/2)\right)^2}{(\wp(z)-\wp(1/2)) (\wp(z)-\wp(1/2+\tau/2))} \cr
Y_{\pm \frac14,0} &=& \frac{1}{128 K^3 (1-k^2)^2} \cr
Y_{\frac12\tau+\pm \frac14,0} &=& \frac{k^2}{128 K^3 (1-k^2)^2} \cr
\eea
We have
\beq
Y_{a,k}  =  \sum_{j=0}^k (\delta_{k-j,1}+(\wp(a)-\wp(\tau/2))U_{a,k-j}) (\delta_{j,0}+(\wp(a+\tau/2)-\wp(\tau/2))\td U_{a,j}) 
\eeq
where
\bea
U_{a,k} &=& \sum_{l=0}^k (-1)^l (\wp''(a))^{-l-1} \sum_{k_1+\dots+k_l = k, \ k_i\geq 1} \prod_{i=1}^l Q_{k_i+1}(\wp(a)) \cr
\td U_{a,k} &=& \sum_{l=0}^k (-1)^l (\wp(a)-\wp(a+\tau/2))^{-l-1} \sum_{k_1+\dots+k_l = k, \ k_i\geq 1} \prod_{i=1}^l Q_{k_i}(\wp(a)) .
\eea

\subsection{Painlev\'e 6, Schlessinger, Liouville CFT}

Let $Z_i$, $i=1\dots 4$, four distinct complex points, and $\alpha_i, i=1,\dots,4$ four complex values called "charges".
We define $\psi(x) = \prod_{i=1}^4 (x-Z_i)$.

Our goal is to consider an elliptic spectral curve of degree 2 ($\deg x=2$) and such that $ydx$ has simple poles over the preimages $(\zeta_{z_i},-\zeta_{z_i}) = x^{-1}(Z_i)$ with residues $\pm \alpha_i$ and no other pole.

This implies that the spectral curve's equation must take the form
\bea
y^2 
&=& \frac{1}{\psi(x)} \left( H + \sum_{i=1}^4 \frac{\alpha_i^2 \psi'(Z_i)}{x-Z_i}  \right) = \frac{P_4(x)}{\prod_{i=1}^4 (x-Z_i)^2} .
\eea
where the arbitrary constant $H\in \CC$ is called the auxiliary parameter.

The numerator is a polynomial of degree 4
\bea
P_4(x) &=& H\psi(x) + \sum_{i=1}^4 \alpha_i^2 \psi'(Z_i) \frac{\psi(x)}{x-Z_i} .
\eea
The underlying curve has genus 1, it is a Torus of some modulus $\tau$.
The modulus $\tau$ ia a function of the zeros of $P_4$.

The spectral curve can be parametrized as follows:
\beq
\begin{cases}
\curve &= \CC/\ZZ+\tau\ZZ \cr
x &= \frac{\beta}{\wp(z)-\wp(\zeta_\infty)} + \alpha \cr
y &= -\frac{C}{\beta^2} \ \frac{\wp'(z) \ (\wp(z)-\wp(\zeta_\infty))^2 }{\prod_{i=1}^4 (\wp(z)-\wp(\zeta_{Z_i}))}  \cr
B(z_1,z_2)  &=  (\wp(z_1-z_2)+G_2) dz_1 \otimes dz_2 \cr
\end{cases}
\eeq
\bea
\text{where } && \alpha = \frac{\wp(\zeta_1)-\wp(\zeta_\infty)}{\wp(\zeta_1)-\wp(\zeta_0)} 
\quad , \quad \beta = \alpha(\wp(\zeta_\infty)-\wp(\zeta_0)) \cr
&& C = \frac{1}{\wp'(\zeta_\infty)} \sqrt{H\prod_{i=1}^4 (\wp(\zeta_\infty)-\wp(\zeta_{Z_i}))}  \cr
\eea
where $x(\pm \zeta_Z) = Z$ for each $Z=0,1,\infty, Z_1,Z_2,Z_3,Z_4$.

It has 4 ramification points
\bea
R = \{0,1/2,\tau/2,1/2+\tau/2\} = \frac12 (\ZZ+\tau \ZZ).
\eea
For each of them the involution is
\beq
z\mapsto \sigma_a(z)=-z.
\eeq
We have
\bea
ydx &=& \frac{C}{\beta} \ \frac{\wp'(z)^2}{\prod_{i=1}^4 (\wp(z)-\wp(\zeta_{Z_i}))} \ dz .
\eea
Therefore, for each $a\in R$, we have
\bea
\frac{1}{dz (z-a)^2}\sum_{k=0}^\infty Y_{a,k} (z-a)^{2k} 
&=& \frac{1}{2ydx} \cr
&=& \frac{\beta}{2C dz} \ \frac{\prod_{i=1}^4 (\wp(z)-\wp(\zeta_{Z_i}))}{\wp'(z)^2} \cr
\sum_{k=0}^\infty Y_{a,k} (z-a)^{2k} 
&=& \frac{\beta}{2C} \frac{(z-a)^2}{\wp'(z)^2} \  \prod_{i=1}^4 (\wp(z)-\wp(\zeta_{Z_i})) .
\eea
We thus obtain
\bea
Y_{a,k}
&=& \frac{\beta}{C} \sum_{k_0+k_1+k_2+k_3+k_4=k} Y^{\text{Weierstrass}}_{a,k_0} \ \prod_{i=1}^4 (Q_{k_i}(\wp(a)) - \delta_{k_i,0} \wp(\zeta_{Z_i}))
\eea
where $Y^{\text{Weierstrass}}_{a,k}$ is given in equation \eqref{eq:YakforWeierstrass}.


\section{Newton's polygon}

Consider an algebraic plane curve over a subfield $\Field \subset \CC$, of equation
\beq
0 = P(x,y) = \sum_{(i,j)\in \Newt} P_{i,j} x^i y^j
\qquad , \quad P_{i,j}\in \Field .
\eeq
Let
\beq
\Delta(x) = \operatorname{Resultant}(P(x,.),P_y(x,.)).
\eeq
The ramification points $a=(x_a,y_a)$ are the common zeros of $P(x,y)$ and $P_y(x,y)$, and not zeros of $P_x(x,y)$.
$x_a$ is thus a zero of $\Delta(x)$.
$\Delta(x)$ is proportional to $\prod_{a\in R}(x-x_a)$ and thus has double zeros at each $a$, whereas $P_y(x,y)$ has simple zeros at $a$. This implies that the ratio $\Delta(x)/P_y^2$ is finite at ramification points. Its only poles can be at $x$ or $y$ infinite, and thus it must be a polynomial of $x,y$:
\beq
\frac{\Delta(x)}{P_y(x,y)} = U(x,y) \in \Field[x,y].
\eeq
Consider the Field extension:
\beq
\Field_\Delta = \Field[\text{zeros of }\Delta].
\eeq
We have for all branchpoints 
\beq
x_a \in \Field_\Delta 
\quad , \quad 
y_a \in \Field[x_a] \subset \Field_\Delta .
\eeq
Let us expand near $a$
\beq
\begin{cases}
x=x_a- \frac{P_{yy}(a)}{2P_x(a)} \zeta_a^2  \cr
y = y_a+ \zeta_a + \sum_{k\geq 2} y_{a,k} \zeta_a^k
\qquad \ y_{a,k} \in \Field[x_a]
\end{cases}
\eeq
The coefficients $y_{a,k}$ are determined by solving $P(x,y)=0$ to each order in $\zeta$.

$c_a = (-P_{yy}(a)/2P_x(a))$. For $k\geq 2$ we have the recursion
\bea
 P_{yy}(a) y_{a,k} 
&=& - [\zeta_a^{k+1}] \sum_{3\leq 2p+q\leq k+1} \frac{1}{p! q!}P_{x^p,y^q}(a) c_a^p \zeta_a^{2p+q} \left(1+\sum_{j\geq 2}y_{a,j}\zeta_a^{j-1}\right)^q \cr
&=& - [\zeta_a^{k+1}] \sum_{3\leq 2p+q\leq k+1} \frac{1}{p! q!}P_{x^p,y^q}(a) c_a^p \zeta_a^{2p+q} \sum_{l=0}^q \binomial{q}{l} \left(\sum_{j\geq 1}y_{a,j+1}\zeta_a^{j}\right)^l \cr 
&=& -  \sum_{3\leq 2p+q\leq k+1} \frac{c_a^p}{p! q!}P_{x^p,y^q}(a)   \sum_{l=0}^q \binomial{q}{l} \sum_{k_1+\dots+k_l = k+1-2p-q} \prod_{j=1}^l y_{a,j+1} .
\eea

$\bullet$ \textbf{Kernel $B$:} 
The kernel $B$ is the following $1\otimes 1$ form (see \cite{EynB})
\beq\label{defBarithm}
B_0((x,y);(x',y')) =  \ \frac{-\frac{P(x,y')P(x',y)}{(x-x')^2(y-y')^2} + Q(x,y;x',y') + S(x,y;x',y')}{P_y(x,y) P_y(x',y')} \ dx \ dx' 
\eeq
where $Q\in \Field[x,y,x',y'] $ is a polynomial
\bea\label{eq:QforB}
Q(x,y;x',y')
&=& \sum_{(i,j)\in \mathcal N} \sum_{(i',j')\in \mathcal N} P_{i,j} P_{i',j'}
\sum_{(u,v)\in \ZZ^2\cap \text{ triangle }(i,j),(i',j'),(i,j')}
|u-i| \ | v-j'| \cr
&& \Big( \delta_{(u,v)\notin {\stackrel{\circ}{\mathcal N}} \cup [(i,j),(i',j')] } \ \ x^{u-1}y^{v-1} x'^{i+i'-u-1}y'^{j+j'-v-1} \cr
&& +\delta_{(u,v)\notin {\stackrel{\circ}{\mathcal N}} \text{ and } (i+i'-u,j+j'-v)\in {\stackrel{\circ}{\mathcal N}} } \ \ x'^{u-1}y'^{v-1} x^{i+i'-u-1}y^{j+j'-v-1}  \cr
&& + \frac12\ \delta_{(u,v)\in  [(i,j),(i',j')]} \ \ x^{u-1}y^{v-1} x'^{i+i'-u-1}y'^{j+j'-v-1}  \Big) \ .
\eea
and $S\in \Field[x,y,x',y'] $ is an arbitrary  polynomial with $(x,y)\Longleftrightarrow(x',y')$ symmetry, and with coefficients only inside the interior ${\stackrel{\circ}{\mathcal N}}$:
\bea
S(x,y;x',y') &=& \sum_{(i,j)\in {\stackrel{\circ}{\mathcal N}}, (i',j')\in{\stackrel{\circ}{\mathcal N}}} S_{(i,j),(i',j')} x^{-i-1}y^{j-1}x'^{-i'-1}y'^{j'-1}.
\eea




$\bullet$ \textbf{Basis :}
We define
\bea
d\xi_{a,0} &=& \Big( - \frac{P(x,y_a) P(x_a,y) }{(x-x_a)^2(y-y_a)^2}  + Q(x_a,y_a;x,y) \cr 
&& + S(x_a,y_a;x,y) \Big) \frac{2 c_a}{P_{yy}(a)} \frac{dx}{P_y(x,y)} \cr
d\xi_{a,k+1} &=& 
d(d\xi_{a,k}/dx) +  \sum_{b\neq a} C_{b,a,k} d\xi_{b,0} 
\eea
where the coefficients $C_{b,a,k}$ are chosen such that $d\xi_{a,k}$ is meromorphic on $\curve$, it has a pole at $z=a$ and no other poles (in particular no pole at ramification points $b\neq a$), and behaves like
\bea
d\xi_{a,k} 
&\sim_a &  (-2)^{-k} (2k+1)!! \ \frac{ d\zeta_a}{ \zeta_a^{2k+2}} + \text{holomorphic} \cr
\eea

\bea
d\xi_{a,k} &=& \frac{R_{a,k}(x,y)}{(x-x_a)^{k+1}} \ \frac{  dx}{P_y(x,y)} \cr
R_{a,0} &=& -\left( \sum_{j=1}^{\deg_x P} \frac{1}{j!} (x-x_a)^{j-1} P_{x^j}(a) \right)\left( \sum_{j=2}^{\deg_y P} \frac{1}{j!} (y-y_a)^{j-2} P_{y^j}(a) \right) \cr 
&& + (x-x_a)\left(Q(a;x,y) + S(a;x,y) \right)  
\qquad \in  \Field_\Delta[x,y] \cr
R_{a,k+1} 
 &=& -(k+1) R_{a,k} + (x-x_a) \partial_x R_{a,k} \cr 
&& - \frac{U}{\Delta'(x_a)}\left(  P_x P_y \partial_y R_{a,k} + R_{a,k} P_y P_{x,y} - R_{a,k} P_x P_{y,y} \right) \cr
 && - \sum_{b\neq a} \frac{(x-x_a)}{\Delta'(x_b) (x-x_b)}\Big( U P_x P_y \partial_y R_{a,k} + U R_{a,k} P_y P_{x,y} \cr 
 && - U R_{a,k} P_x P_{y,y} -2 U(b) R_{a,k}(b)  \frac{(x-x_a)^k}{(x_b-x_a)^k} R_{b,0} \Big) \cr
&& \in \Field_\Delta[x,y] .
\eea

We have
\bea
B_{\text{odd}}(x,y;x',y') 
&=& - \sum_k \frac{(-2)^k}{(2k-1)!!}  \zeta_a^{2k}d\zeta_a \ \otimes \ d\xi_{a,k}(x',z')  .
\eea

\subsubsection{Example: Cubic elliptic curve}

Let $t$ be a formal variable and $\Field =\mathbb Q[t]$. Consider the algebraic plane curve 
\bea
P(x,y) &=& x^3 + y^3 + t xy +1 .
\eea
Its discriminant is
\bea
\Delta(x) = 27 (x^3+1)^2 + 4 t^3 x^3 .
\eea
We have
\bea
P_y(x,y) &=& 3 y^2 + tx \cr
U(x,y) &=& \frac{\Delta(x)}{P_y(x,y)^2} = 3y^2 + 4 tx.
\eea

The 6 ramification points $a=(x_a,y_a)$ are the 6 zeros of $\Delta(x)$. They can be written (6 = 3 possible cubic roots and 2 possible square roots):
\bea
y_a &=& \frac{-t}{3} \left( 1 \pm \sqrt{1+27/t^3} \right)^{\frac13} \cr
x_a &=& \frac{-3}{t} y_a^2 = \frac{-t}{3} \left( 1 \pm \sqrt{1+27/t^3} \right)^{\frac23}  .
\eea

For the kernel $B$, equation \eqref{eq:QforB} gives
\bea
Q(x,y'x',y') &=& xy+ 2xy'+2x'y+x'y' \cr
S(x,y'x',y') &=& S .
\eea

This gives
\bea
d\xi_{a,0} &=& \frac{R_{a,0}}{(x-x_a)} \ \frac{dx}{P_y} \cr
R_{a,0} 
&=& - (y+2y_a)(x^2+x x_a + x_a^2 + t y_a) \cr 
&& + (x-x_a)(xy+2x y_a+2y x_a + x_a y_a +S) \cr
&=& (x-x_a)(S-3 x_a y_a) - (y+2y_a)(3x_a^2+ty_a) \cr
&=& (x-x_a)(S-3 x_a y_a) - (y-y_a)(3x_a^2+ty_a) - 3y_a (3x_a^2+t y_a)\cr
\eea


\section{Higher order ramification points}
\label{sec:higherbp}

Let us now consider spectral curves whose ramification points can be of higher order, let
\beq
r = \max\{ \text{order of ramification points} \}.
\eeq
Regular spectral curves have $r=2$.
Here we consider arbitrary $r\geq 2$, and use the construction of \cite{BElocalglobal}.

\subsection{TR and tensors}

For $z\in\curve$ generic, 
define $\tau'_n(z)$ the set of all possible $n-$uples in $x^{-1}(z)\setminus\{z\}$. $\tau'_n(z)=\emptyset$ if $n+1>\# x^{-1}(z)=\deg x$.
For $n\geq  3$, let $z_2\in\curve$ and $(z_3,\dots,z_n) \in \tau'_{n-2}(z_2)$,  define the kernel
\beq
K^{(n)}_a(z_1;z_2, \overbrace{z_3,\dots,z_n}^{\in \tau'_{n-2}(z_2)}  )
=  \frac{\int_{a}^{z_2} B(z_1,.)}{\prod_{j=3}^{n} (y(z_2)-y(z_j))dx(z_2)} .
\eeq

For $n\geq 1$ and $n+m+2h \geq 3$, we define the tensor $A^{(h)}_{n|m} \in V^{\otimes n} \otimes V^{*\otimes m}$ by:
\bea
&& \sum_{\alpha_1,\dots,\alpha_n}
A^{(h)}_{n|m}[\alpha_1,\dots,\alpha_n |  \alpha_{n+1},\dots, \alpha_{n+m}  ]  \ d\xi_{\alpha_1}(z_1) d\xi_{\alpha_2}(\td z_2)  \dots d\xi_{\alpha_n}(\td z_n) \cr
&=&  \sum_{a\in R} \Res_{z_2\to a} \sum_{(z_3,\dots,z_{n+m+2h}) \in \tau'_{n+m+2h-2}(z_2)}  K^{(n+m+2h)}_a(z_1;z_2,\dots, z_{n+m+2h} ) \cr 
&& \prod_{j=2}^n B(z_j,\td z_j) \prod_{j=n+1}^{n+m} d\xi_{\alpha_j}(z_j)  \prod_{j=1}^{h} \frac{B(z_{n+m+2j-1},z_{n+m+2j})}{2^{\delta(n+m+2j-1>2)}}  .
\eea
If $n<1$ or $n+m+2h<3$ we define $A^{(h)}_{n|m}=0$.

In fact, it was shown in \cite{BElocalglobal} that for a given ramification point $a$ of order $r_a$, its contribution to $A^{(h)}_{n|m}$ vanishes if $n+m+2h>r_a+1$.
Therefore if $r=\max(\{ r_a\}_{a\in R}\}$, we need all tensors such that $n+m+2h \leq r+1$.

\smallskip
For simple ramification points $r=2$, the 4 possible tensors with $n\geq 1$ and $3\leq n+m+2h \leq r+1=3$, i.e. $n+m+2h=3$, are the tensors $A,B,C,D$:
\beq
\begin{cases}
A =& A^{(0)}_{3|0} \cr
B =& A^{(0)}_{2|1} \cr
C =& A^{(0)}_{1|2} \cr
D =& A^{(1)}_{1|0} \cr
\end{cases}
\eeq
The Topological Recursion of \cite{BElocalglobal} amounts to, for $2g-2+n>0$ and $n\geq 1$:
\bea
 F_{g,n}[\alpha_1,\dots,\alpha_n] 
&=& \sum_{h=0}^{\lfloor  r/2 \rfloor} \ \  \sum_{p=0}^{r-2h}  \sum_{\td\alpha_1,\dots,\td\alpha_p} \cr
&& \sum_{\mu \vdash \{ \td\alpha_1,\dots,\td\alpha_{p}\} }  \ \ 
\sum_{I_1\sqcup \dots \sqcup I_{\ell(\mu)} = \{\alpha_{2},\dots,\alpha_n\}} \ \ 
\sum_{g_1+\dots+g_{\ell(\mu)}=g+\ell(\mu)-p} \cr
&& \prod_{i=1}^{\ell(\mu)} \delta(2g_i-2+|\mu_i|+|I_i|\geq 0) \cr
&& \delta(U = \cup_{i, \ g_i=0, |\mu_i|=1, |I_i|=1} \ I_i )  \delta(\overline{U} = \{\td\alpha_1,\dots,\td\alpha_p\}\setminus  U)
\cr
&& \delta(h=\# \{i, \ g_i=0, |\mu_i|=2, |I_i|=0 \}  ) \cr 
&& A^{(h)}_{1+|U|,p-|U| }\left[\alpha_1, U| \overline{U} \right]  \prod_{i, \ 2g_i-2+|\mu_i|+|I_i|>0}  F_{g_i,|\mu_i|+|I_i|}[\mu_i,I_i]  .
\eea

\subsection{$(r,s)$ curves and r-spin}

Let $r\geq 2$, and $s> 1-r$, and $s\neq 0 \mod r$, and as in \cite{BBNRSrs25}, consider the $(r,s)$ spectral curve:
\beq
\begin{cases}
x(z) = \frac{1}{r} z^r \cr
y(z) = z^s \cr
B(z_1,z_2) = \frac{1}{(z_1-z_2)^2} dz_1 \otimes dz_2
\end{cases}
\eeq
There is one ramification point $R=\{0\}$, of order $r$.
Let 
$$
\rho = e^{2i\pi/r}.
$$

Define the basis:
\bea
d\xi_{\alpha}(z) = z^{-\alpha-1}dz.
\eea
We have
\bea
B(z_1,z_2) &\sim_{z_2\to 0} &  \sum_{\alpha\geq 1}  \alpha z_2^{\alpha-1}dz_2 \ d\xi_\alpha(z_1) \cr
dS(z_1,z_2) = \int_0^{z_2} B(z_1,.) &\sim_{z_2\to 0} & \sum_{\alpha\geq 1}   z_2^{\alpha} \ d\xi_\alpha(z_1) .
\eea

Kernel:
\bea
K^{(n)}(z_1;z_2,\dots,z_n)
&=& \frac{ \sum_{\alpha_1\geq 1} z_2^{\alpha_1} d\xi_{\alpha_1}(z_1)}{\left( z_2^{r-1}dz_2\right)^{n-2}} \frac{1}{\prod_{j=3}^n (z_2^{s}-z_j^{s})} \cr
&=& \sum_{\alpha_1\geq 1} d\xi_{\alpha_1}(z_1) \frac{  z_2^{\alpha_1-(n-2)(s+r-1)} }{\left( dz_2\right)^{n-2} \ \prod_{j=3}^n (1-(z_j/z_2)^{s})} .
\eea

Let us define $\theta_j=1$ if $j\leq n$ and $\theta_j=-1$ if $j>n$.
The operator $A^{(h)}_{n|m}$ is then equal to
\bea
&& A^{(h)}_{n|m}[\alpha_1,\dots,\alpha_n |  \alpha_{n+1},\dots, \alpha_{n+m}  ] \cr
&=& \Res_{z_2\to 0} \frac{  z_2^{\alpha_1-(n+m+2h-2)(s+r-1)} }{\left( dz_2\right)^{n+m+2h-2}} \ \ \sum_{\sigma\in \mathfrak S_{r+1}, \ \sigma_1=r+1,\sigma_2=r}    \frac{1}{\prod_{j=3}^{n+m+2h} (1-\rho^{s\sigma_j })}  \cr 
&& \prod_{j=n+1}^{n+m} \rho^{-\alpha_j \sigma_j} z_2^{-\alpha_j-1}dz \prod_{j=2}^n \alpha_j \rho^{\alpha_j \sigma_j} z_2^{\alpha_j-1} dz_2 \cr
&& \prod_{j=1}^{h} 2^{-\delta(n+m+2j-1>2)} \frac{\rho^{\sigma_{n+m+2j-1}}\rho^{\sigma_{n+m+2j}}}{(\rho^{\sigma_{n+m+2j-1}}-\rho^{\sigma_{n+m+2j}})^2}  \frac{dz_2^2}{z_2^2}  \cr
&=& \frac{2^{\delta_{n+m,1 \text{ and } h>0}}}{2^h}\prod_{j=2}^n \alpha_j \Res_{z\to 0} dz \ \   z^{\sum_{j=1}^{n+m} \theta_j \alpha_j - (n+m+2h-1) -(n+m+2h-2)(s+r-1) } \ \cr 
&& \sum_{\sigma\in \mathfrak S_{r+1}, \ \sigma_1=r+1,\sigma_2=r}   \    \  \frac{\prod_{j=2}^{n+m} \rho^{\alpha_j \sigma_j \theta_j}}{\prod_{j=3}^{n+m+2h} (1-\rho^{s\sigma_j })}   \prod_{j=1}^{h}  \frac{\rho^{\sigma_{n+m+2j-1}}\rho^{\sigma_{n+m+2j}}}{(\rho^{\sigma_{n+m+2j-1}}-\rho^{\sigma_{n+m+2j}})^2} \cr
&=& \prod_{j=2}^n \alpha_j \ \ \delta\left(\sum_{j=1}^{n+m} \theta_j \alpha_j = (n+m+2h-2)(s+r) \right)  \cr 
&& \qquad C^{(h)}_{n+m}(\theta_2\alpha_2,\dots,\theta_{n+m}\alpha_{n+m}) \cr
\eea
where
\bea
 C^{(h)}_{n}(\alpha_2,\dots,\alpha_{n}) 
&=& \frac{2^{\delta_{n+m,1 \text{ and } h>0}}}{2^h} \sum_{\sigma\in \mathfrak S_{r+1}, \ \sigma_1=r+1,\sigma_2=r}   \    \  \frac{\prod_{j=2}^{n} \rho^{\alpha_j \sigma_j }}{\prod_{j=3}^{n+2h} (1-\rho^{s\sigma_j })}  \cr 
&& \qquad \quad \prod_{j=1}^{h}  \frac{\rho^{\sigma_{n+2j-1}}\rho^{\sigma_{n+2j}}}{(\rho^{\sigma_{n+2j-1}}-\rho^{\sigma_{n+2j}})^2} .
\eea


$\bullet$ Example $r=2,s=\text{odd}$, we have $\rho=-1$. 
The only operators have $n+m+2h=3$:
We have
\bea
 C^{(0)}_{3}(\alpha_2,\alpha_{3}) 
&=& \frac{(-1)^{\alpha_3}}{1-(-1)^s} = \frac{(-1)^{\alpha_3}}{2}.
\eea
\bea
 C^{(1)}_{1}(\emptyset) 
&=& \frac{1}{1-(-1)^s} \frac{-1}{(1-(-1))^2} = \frac{-1}{8}.
\eea
Writting $\alpha_j=2d_j+1$ this gives
\bea
A^{(0)}_{3|0}[2d_1+1,2d_2+1,2d_3+1]
&=& \frac{-1}{2} (2d_2+1)(2d_3+1)\delta(d_1+d_2+d_3=\frac12(s-1)) \cr
A^{(0)}_{2|1}[2d_1+1,2d_2+1|2d_3+1]
&=& \frac{-1}{2} (2d_2+1) \delta(d_1+d_2= d_3+\frac12(s+1))  \cr
A^{(0)}_{1|2}[2d_1+1|2d_2+1,2d_3+1]
&=& \frac{-1}{2} \delta(d_1= d_3+d_2+\frac12(s+2))  \cr
A^{(1)}_{1|0}[2d_1+1]
&=& \frac{-1}{8} \delta(d_1= 1)  .
\eea

$\bullet$ Example $r=3$ and $s\neq 0 \mod 3$, we have $\rho=e^{2i\pi/3}$.
Remark that $(1-\rho^s)(1-\rho^{-s}) = 3$ and $(\rho-\rho^{-1})^2 = -3$.
The only possibilities are such that $n+m+2h=3,4$, which correspond to 11 possible tensors.

We have
\bea
C_4^{(0)}(\alpha_2,\alpha_3,\alpha_4)
&=& \frac{\rho^{\alpha_3-\alpha_4}+\rho^{\alpha_4-\alpha_3}}{(1-\rho^s)(1-\rho^{-s})} = \frac13 (\rho^{\alpha_3-\alpha_4}+\rho^{\alpha_4-\alpha_3})  \cr
C_2^{(1)}(\alpha_2)
&=& 2 \frac13 \frac{\rho^{1+2}}{2(\rho-\rho^{-1})^2} = \frac{-1}{9} \cr
C_3^{(0)}(\alpha_2,\alpha_3)
&=& \frac{\rho^{\alpha_3}}{1-\rho^s}+\frac{\rho^{-\alpha_3}}{1-\rho^{-s}} = \frac{\rho^{\alpha_3}-\rho^{s-\alpha_3}}{1-\rho^s} \cr
C_1^{(1)}(\emptyset)
&=& \sum_{\sigma_3=1}^2 \frac{\rho^{\sigma_3}}{(1-\rho^{s\sigma_3})(1-\rho^{\sigma_3})^2} = \frac{-1}{3}
\eea
and
\bea
&& A^{(h)}_{n|m}[\alpha_1,\dots,\alpha_n |  \alpha_{n+1},\dots, \alpha_{n+m}  ] \cr
&=& \prod_{j=2}^n \alpha_j \ \ \delta\left(\sum_{j=1}^{n+m} \theta_j \alpha_j = (n+m+2h-2)(s+r) \right)  \cr 
&& \qquad C^{(h)}_{n+m}(\theta_2\alpha_2,\dots,\theta_{n+m}\alpha_{n+m}) \cr
\eea


\section{Algorithmic implementation}

\subsection{Issues with standard tensor libraries}

The implementation of TR is mostly about tensor products and contractions.
There exists many tensor libraries, often implemented using GPU, in python and other languages, however, we found several issues to use them:

\begin{itemize}
    \item Tensor contractions are over indices that can appear in different permuted orderings. In other words we need tensors whose indices are not labeled by [0,1,2,3...,rank] but by some labels which are not always in the same order.
    Moreover in the recursive procedure, the same tensor $F_{g,n}$ is re-used many times, with different labellings.
    The idea is to store tensors only once, don't permute their columns, but only permute the labelling, which is less time consuming.

    \item The rank $n$ tensors $F_{g,n}$ have finite dimension.
    But the tensors $A,B,C,D$ and $A^{(h)}_{n|m}$ have typically infinite dimension. 
    For example when applying the tensor $B$, we need to perform the operation
    \beq
    \sum_{\beta} B[\alpha_1,\alpha_j|\beta] \  F_{g,n-1}[\alpha_2,\dots,\alpha_{j-1},\beta,\alpha_{j+1},\dots,\alpha_n]
    \eeq
    For $B[\alpha_1,\alpha_j|\beta]$ in principle there can be an infinite number of values of $\alpha_1,\alpha_j,\beta$.
    However since the tensor $F_{g,n-1}$ is of finite dimension, there are finitely many possible values of $\alpha_2,\dots,\alpha_{j-1},\beta,\alpha_{j+1},\dots,\alpha_n$, in particular of $\beta$.
    Then, remark that for a given $\beta$, there are only finitely many values of $\alpha_1,\alpha_j$ for which  $B[\alpha_1,\alpha_j|\beta]\neq 0$, so that eventually, the sum is a finite sum, and the result is a finite dim tensor.
    
    Therefore we need a tensor library that contains the information that for given out-indices, only finitely many possible in-indices are attainable.
    And moreover since the tensors $A^{(h)}_{n|m}$ have infinitely many possible indices, the value can't be stored as a tensor, but as functions of indices.

    \item A less critical issue, is the fact that the tensors $F_{g,n}$ are symmetric. Storing all the values is very redundant ($n!$ times too much), and could be optimized. 

\end{itemize}

\subsection{Implementation}

For these reasons, I chose to define a new Tensor implementation (in python) with the following classes: 

\medskip
$\bullet$ \textbf{Indices} is a class for list of indices $I=[\alpha_1,\alpha_2,\dots,\alpha_{\dim}]$, where the indices $\alpha_i$ are not necessarily integers, they are typically pairs $\alpha_i=(a_i,d_i)=(point,integer)$, in fact they can be any objects.

What we need is that Indices-objects are sortable (and thus have comparison methods $\leq, \geq, =, \neq$), and iterable, and one can access their $i^{th}$ elements by a \texttt{\_\_getitem\_\_} method. 
The class must be like that:

\smallskip
\begin{tabular}{|c|c|}
\hline
    Class    & \textbf{Indices}  \\
\hline 
    Attributes & list [index${}_0$,...,index${}_{n-1}$]  \\ 
\hline 
    Methods 
      & $\leq, \geq, =, \neq$  \\ 
      & \texttt{dim} = length  \\ 
      & \texttt{sort}  \\ 
      & \texttt{\_\_getitem\_\_} \\
      & \texttt{\_\_iter\_\_} \\
      & \texttt{\_\_hash\_\_} \\
\hline 
\end{tabular}

\smallskip
\textbf{Example:} $I=[\alpha_0,\dots,\alpha_{n-1}]$, $I[1]=\alpha_1$, $I_1<I_2$,..., \\
\texttt{for i in I: do something}

\medskip
$\bullet$ \textbf{TRTensor} is a class for Tensors. This is the class to store the $F_{g,n}$s.
A Tensor $T$ is a dictionary whose keys are elements of the class Indices, and whose values are numerical values. It must have a \texttt{\_\_getitem\_\_} method such that $T[I]=$value.
It must have also linear operations: addition and scalar multiplication.
It must also have a tensor product $\otimes$ and a contraction (dot) method.
For contractions and tensor products, it is useful to contract on labels rather than column numbers. Therefore each Tensor has a labelling of columns, and the dot and tensor method allow optional relabelling before contracting.

\smallskip
\begin{tabular}{|c|c| l |}
\hline
    Class    & \textbf{TRTensor}  & \\
\hline 
    Attributes & dict of \{Indices : value\} &  \\ 
     & labels = dict {name:rank} &  \\ 
\hline 
    Methods 
      & \texttt{\_\_getitem\_\_} & \texttt{A[I]} \\
      & $+,-$ & \texttt{A+B} \\ 
      & $*$scalar &  \texttt{A*x} \\ 
      & \texttt{dot}  $.$ & \texttt{A.dot(B)} \ : \  contract over common labels  \\ 
      & \texttt{tensor} $\otimes$ & \texttt{A.tensor(B)} \\ 
\hline 
\end{tabular}

\medskip

\textbf{Example:}
Let I${}_1$=Indices([1,7,2]), I${}_2$=Indices([0,3,1]) \\
T=Tensor(\{I${}_1$:7,I${}_2$:3\},labels=\{ "d1":0,"d2":1,"d3":2 \})  \\

\bigskip
$\bullet$ \textbf{TROperator} is a class for the tensors $A^{(h)}_{n|m}$ (and we recall that $A=A^{(0)}_{3|0}, B=A^{(0)}_{2|1}, C=A^{(0)}_{1|2}, D=A^{(1)}_{1|0} $.
It can't be a dictionnary, because it may have an infinite number of pairs \{key:value\}. Co
Moreover, the keys are list of Indices (in the class $\textbf{Indices})$, and they are in fact of two types: in-Indices (dim= $n$)  and out-indices (dim = $m$).
For a given out-Indices $I$ of dim $m$, we must have a function that returns the list of all possible in-Indices, so a function Indices$\mapsto$[list of Indices].
And then for each pair (in-Indices,out-Indices) we must have a function that returns the value. This function can be used ti overload the \texttt{\_\_getitem\_\_} method.
Then, we must have a contraction (dot) method that contracts a \textbf{TROperator} with a \textbf{TRTensor}, again on labels rather than column numbers. Therefore each \textbf{TROperator} has a labelling of columns, and the dot method allows optional relabelling before contracting.

\smallskip

\begin{tabular}{|c|c| }
\hline
    Class    & \textbf{TROperator}   \\
\hline 
    Attributes & n,m,h  \\
     & labels\_in , labels\_out  \\
    & function \texttt{\_outindicesfromin}(out-Indices)   \\ 
    &  $\mapsto$ [list of possible in-Indices]   \\ 
     & function \texttt{\_functionindices}(out-Indices,in-Indices) $\mapsto$ value   \\ 
\hline 
    Methods & \texttt{\_\_getitem\_\_}  \\
      & \texttt{dot}  $.$  \texttt{A.dot(B)} $\mapsto$ TRTensor\ \\ 
      & \  contract over common labels  \\ 
\hline 
\end{tabular}

\smallskip
\textbf{Example:} \\
f\_outindicesfromin($d_2$) = $\{(d0,d1) \ | d_0+d_1= d_2+1 \}$ \\
f\_functionindices($d_0,d_1,d_2$) = $ -\frac{(2d_2+1)!!}{4 (2d_0+1)!!(2d_1-1)!!}  \ \delta_{d_0+d_1, d_2+1} $ \\
B = TROperator(2,1, f\_outindicesfromin,f\_functionindices, labels\_in = \{ "d1":0,"d4":1\}, labels\_out = \{"beta":2 \}  ) \\

B.dot( Tensor(labels=\{ "d2":0,"d3":1,"beta":2,"d5":3 \}) ) the contraction will be over the only common label "beta". 
The result is a tensor of rank 5: Tensor(labels=\{ "d1","d2","d3","d4","d5" \})

\bigskip
$\bullet$ \textbf{TRQAS} is the class for Quantum Airy Structures. It contains 2 dictionaries: the dictionary to store $A,B,C,D$ and more generally $A^{(h)}_{n|m}$:  \{(n,m,h):TROperator\}, and a dictionary to store (cache) the $F_{g,n}$: \{(g,n):TRTensor\}.
It also contains a function Xi, that for an index $\alpha$ returns the differential form $d\xi_\alpha$ (as a Lambda function of $z$), in order to be able to compute the $W_{g,n}$.
It may also contain more methods useful for Topological Recursion...

\smallskip
\begin{tabular}{|c|c| }
\hline
    Class    & \textbf{TRQAS} {\it (QAS = Quantum Airy Structure)}   \\
\hline 
    Attributes 
     & dict \{ (g,n) : TRTensor \}  \\
     & dict \{ (n,m,h) : TROperator \}  \\
\hline 
    Methods & \_compute(g,n)  \\
     & function Xi(d) $\mapsto$ (function(z)$\mapsto$ value)  \\
\hline 
\end{tabular}

\bigskip
$\bullet$ I also implemented more classes.
The purpose is to transform a spectral curve data (typically $\curve,x,y,B,R$) into Quantum Airy Structure operators.
For example a class \textbf{KdVTimes} and \textbf{SpCurveKdV}, a class \textbf{SpCurveGUE} a class \textbf{Torus} that implements relations \eqref{eq:wpderiveQ}, \eqref{eq:computewpderivQ} and \textbf{SpCurveTorus}, and \textbf{SpCurveNewtonPolygon} and few other examples.

\bigskip
All this is in the gitlab \\
\url{https://gitlab.com/toprec/toprec}
in the \texttt{dev} branch.


\section{Conclusion}

All this has been implemented in a python code.
However we ask motivated readers to help and contribute to improving and developing this implementation.

\section*{Aknowledgements}

The author thanks Dimitrios Mitsios and Vincent Delecroix for discussions, for the gitlab and python and Sagemath package, implementing part of these computations.
This work is supported by the ERC-SyG project, Recursive and Exact New Quantum Theory (ReNewQuantum) which received funding from the European Research Council (ERC) under the European Union's Horizon 2020 research and innovation programme under grant agreement No 810573.

\printbibliography

\end{document}